\documentclass[nohyper,12pt,letterpaper]{JHEP3}
\usepackage{graphicx,epsfig,youngtab}

\author{Warren Carlson$^1$, Robert de Mello Koch$^{1,2}$ and Hai Lin$^{3}$\\
$^{1}$ National Institute for Theoretical Physics,\\
Department of Physics and Centre for Theoretical Physics,\\
University of the Witwatersrand, Wits, 2050, South Africa\\
\qquad\\
$^{2}$Stellenbosch Institute for Advanced Studies,\\
Stellenbosch, South Africa\\
\qquad\\
$^{3}$Department of Particle Physics, Facultad de Fisica,\\
Universidad de Santiago de Compostela, 15782, Santiago de Compostela, Spain\\
\qquad\\
E-mail: \email{Warren.Carlson@students.wits.ac.za, robert@neo.phys.wits.ac.za, hai.lin@usc.es}}

\abstract{ In this article we study operators with a dimension
$\Delta\sim O(N)$ and show that simple analytic expressions for
the action of the dilatation operator can be found. The operators
we consider are restricted Schur polynomials. There are two
distinct classes of operators that we consider: operators labeled
by Young diagrams with two long columns or two long rows. The main
complication in working with restricted Schur polynomials is in
building a projector from a given $S_{n+m}$ irreducible
representation to an $S_n\times S_m$ irreducible representation
(both specified by the labels of the restricted Schur polynomial).
We give an explicit construction of these projectors by
reducing it to the simple problem of addition of angular momentum in
ordinary non-relativistic quantum mechanics. The diagonalization
of the dilatation operator reduces to solving three term recursion
relations. The fact that the recursion relations have only three
terms is a direct consequence of the weak mixing at one loop of
the restricted Schur polynomials. The recursion relations can be
solved exactly in terms of symmetric Kravchuk polynomials or in
terms of Clebsch-Gordan coefficients. This proves that the
dilatation operator reduces to a decoupled set of harmonic
oscillators and therefore it is integrable. }

\preprint{WITS-CTP-065}

\title{Nonplanar Integrability}

\keywords{Giant Gravitons, AdS/CFT correspondence, super Yang-Mills theory}

\def \Tr{\mbox{Tr\,}}

\begin{document}

\section{Introduction}

Ultraviolet violet divergences inevitably appear in perturbative
calculations of various correlation functions in the ${\cal N}=4$
super Yang-Mills theory. These divergences require
renormalization, which induces a mixing among operators with the
same bare conformal dimension. The eigenvalues of the new
eigenstates under conformal rescalings are a sum of the classical
scaling dimension and a loop correction, the anomalous dimension.
By studying the diagrams which contribute to the renormalization
of these operators it is possible to obtain an expression for the
dilatation operator. The eigenvalues of this operator are the
anomalous dimensions. A key discovery has been the understanding
that the dilatation operator can be identified with the
Hamiltonian of an integrable spin chain \cite{Minahan:2002ve}.
This integrability has been found in the planar limit of the
theory.

Integrability has proved to be a key ingredient towards finding
the exact spectrum of composite operators in ${\cal N} = 4$ super
Yang-Mills theory. 
Further, direct perturbative calculations become very cumbersome
at high loop orders but can be avoided, by assuming that the
observed integrability persists to all loop orders. The dilatation
operator can then be determined using general arguments. Clearly,
unraveling the integrable structures in the gauge theory is an
important problem. In the AdS/CFT correspondence
\cite{Maldacena:1997re}, these planar results are directly
relevant to the problem of determining the exact spectrum of
strings in the AdS$_5\times$S$^5$ background.

Summing the planar diagrams gives a valid description of the large
$N$ limit of those operators in the theory with a bare dimension
$\Delta$ such that $\Delta^2/N\ll 1$. To correctly construct the
large $N$ limit for operators in the theory with an even larger
dimension, one has to sum much more than just the planar
diagrams. 
There are many good reasons to study operators with a large bare
dimension. Giant gravitons \cite{McGreevy:2000cw} are 
dual to operators with a dimension of $O(N)$
\cite{Balasubramanian:2001nh,Corley:2001zk,Berenstein:2004kk}
while new background geometries are dual to operators with a
dimension of $O(N^2)$
\cite{Corley:2001zk,Berenstein:2004kk,Lin:2004nb,Balasubramanian:2005mg,Grant:2005qc,Chen:2007gh,shahin,Koch:2008ah,Koch:2008cm,Koch:2009jc,us,hai1,hai2}.
New methods sparked by the pioneering works \cite{Corley:2001zk},
employing group representation techniques, probe the theory beyond
the planar limit. It is now known how to build a basis for these
operators, with the very nice feature that the basis diagonalizes
the two point function
\cite{Corley:2001zk,Balasubramanian:2004nb,de Mello Koch:2007uu,de
Mello
Koch:2007uv,Kimura:2007wy,Bekker:2007ea,Brown:2007xh,Bhattacharyya:2008rb,SS,Brown:2008rr,Kimura:2008wy,Kimura:2009wy,Ramgoolam:2008yr}.
Techniques to compute correlation functions of these large
operators have been developed
\cite{Bhattacharyya:2008rc,Koch:2008ah,Koch:2008cm,Koch:2009jc}
and some investigation into loop corrections have been carried out
\cite{de Mello Koch:2007uu,de Mello
Koch:2007uv,Brown:2008rr,Brown:2008rs,yusuke,Koch:2010gp,Pasukonis:2010rv,VinceKate}.
Formulas for the one loop dilatation operator have been found and
these results show that mixing at one loop is highly constrained.
Much more work is needed. In particular, although formulas for the
one loop dilatation operator are known, they are difficult to
evaluate explicitly. Ultimately we would like both explicit
formulas for the action of the dilatation operator as well as its
eigenvalues and eigenvectors.

Special cases of the problem in the $SU(2)$ sector have been
studied numerically in \cite{Koch:2010gp,VinceKate}. The
computations are rather involved and the expressions for the
dilatation operator are complicated. However, upon solving for the
spectrum of the dilatation operator numerically, one finds a
strikingly simple result: the dilatation operator is equivalent to
a sum of harmonic oscillators. Given this simplicity, one might
hope that a general analytic approach is possible. In this article
we study operators with a dimension $\Delta\sim O(N)$ and show
that simple analytic expressions for the action of the dilatation
operator can be found. Using this construction of the dilatation
operator for this class of operators, we can test whether or not
it is integrable. We will argue that the dilatation operator
reduces to a decoupled set of harmonic oscillators and hence that
it is indeed integrable.

The operators we consider are restricted Schur polynomials built
using $n$ complex matrices $Z$ and $m$ complex matrices $Y$. They
are labeled by three Young diagrams. The first Young diagram
specifies a representation of the symmetric group $S_{n+m}$ and
the second two a representation of $S_n\times S_m$. There are two
distinct classes of operators that we consider: operators labeled
by Young diagrams with two long columns or two long rows. Previous
experience with the half BPS case \cite{Corley:2001zk} suggests
that these operators are dual to sphere giants (long columns) or
AdS giants (long rows). When we say the Young diagrams have two
rows (or columns) we mean that all three Young diagram labels have
two rows (or columns). By saying the columns (or rows) are large
we mean that both of the columns (or rows) of the first Young
diagram label contain $O(N)$ boxes.

The main complication in working with restricted Schur polynomials
is in building a projector from a given $S_{n+m}$ irreducible
representation to an $S_n\times S_m$ irreducible representation
(both specified by the labels of the restricted Schur polynomial).
One of the main technical advances of this article is an explicit
and simple construction of these projectors. We will now outline
the logic of this construction. The projectors organize
(partially) labeled Young diagrams into irreducible
representations of $S_m$. Our first step entails showing that
these partially labeled Young diagrams can be traded for a
(reducible) polynomial representation of the symmetric group. The
advantage of the polynomial representation is that it admits the
action of an operator $d$ introduced by Dunkl in his study of
intertwining functions \cite{Dunkl}. This operator is then used to
construct a Casimir $d^\dagger d$ whose eigenspaces of definite
eigenvalue are precisely the irreducible representations we are
after. Thus, we are able to substitute the problem of constructing
projectors with the eigenproblem of $d^\dagger d$. We then define
one more map, which maps our original partially labeled Young
diagrams into states of a spin chain. When acting on the spin
chain $d^\dagger d$ has a particularly simple form and its
eigenproblem is easily solved explicitly. Indeed, we reduce the
problem of computing projectors to the problem of addition of
angular momentum in ordinary non-relativistic quantum mechanics!
This allows us to give rather explicit and simple formulas for the
action of the dilatation operator. The diagonalization of the
dilatation operator reduces to solving three term recursion
relations. The fact that the recursion relations have only three
terms is a direct consequence of the weak mixing at one loop of
the restricted Schur polynomials. At this point we find that the
dilatation operator is very closely related to certain discrete
models for the harmonic oscillator \cite{finiteoscillator} and as
a consequence the recursion relations can be solved exactly in
terms of symmetric Kravchuk polynomials or in terms of
Clebsch-Gordan coefficients. If we consider the limit in which the
recursion relation can be replaced by a differential equation,
which is the large $N$ limit that we have taken, we find a direct
connection to the usual harmonic oscillator. In this way we claim
that when acting on the class of operators belonging to the
$SU(2)$ sector of ${\cal N}=4$ super Yang-Mills theory and having
a dimension $\Delta\sim O(N)$ the dilatation operator is
integrable.

One of the lessons learned from the study of the half BPS sector
of ${\cal N}=4$ super Yang-Mills theory is that sphere giants and
AdS giants are not independent solutions, but instead are dual to
each other. Indeed, the half BPS states are in one to one
correspondence with Young diagrams of $SU(N)$. A given Young
diagram may be regarded either as labeling a collection of as many
sphere giant gravitons as there are columns or as labeling a
collection of as many AdS giant gravitons as there are rows. This
correspondence can also be transparently seen from the 1/2 BPS
geometries with a white annulus \cite{Lin:2004nb}. For example, a
single AdS giant of angular momentum $k$ may equally well be
thought of as collection of $k$ sphere giant gravitons, each with
unit angular momentum. A similar equivalence works for the full
spectrum of ${1\over 8}$ BPS states: the partition function
obtained by quantizing the sphere giant graviton is the same as
the partition function obtained by quantizing the AdS giant
graviton \cite{Biswas:2006tj,Mandal:2006tk}, see also
\cite{Berenstein:2005aa}. Our computation of the dilatation
operator at one loop uncovers an incredibly simple relation
between the action of the dilatation operator on sphere giants and
its action on AdS giants. Our result suggests that the duality
discovered in the BPS sector may be enlarged to the non-BPS sector
of the theory.

 In the next section
we study the problem in the context of the two sphere giant
system. The new method used to construct the symmetric group
projectors is most easily developed for AdS giants. Setting the
problem up and constructing the projectors is accomplished in
sections 3 and 4 respectively. The explicit expression for the
action of the dilatation operator acting on AdS giants is the last
formula obtained in section 4. Section 5 explains how to build
the construction for sphere giants and uncovers a very simple
relation between the action of the dilatation operator on sphere
giants and its action on AdS giants. In section 6 we consider the
problem of diagonalizing the dilatation operator and obtain the
relevant discrete wave equations. Finally we briefly discuss
interesting features of our results in Section 7.

\section{Two Sphere Giant Gravitons}

{In this section we first review the action of the dilatation
operator on a system of two sphere giant gravitons. The two sphere
giant graviton system is described by restricted Schur polynomials
labeled by Young diagrams that have two long columns. The
expression for the dilatation operator has been evaluated
explicitly in some special cases; a general expression is not
known. This action simplifies considerably in a particular limit
that is described in detail. Given this simplicity, the general
construction of the dilatation operator directly in this limit, is
considered. We argue that the simplicity of this limit can be
exploited when constructing the dilatation operator by employing
Young's orthogonal representation for the symmetric group. We are
able to obtain a general expression for the dilatation operator
acting on the two sphere giant system because we manage to give a
general explicit construction of the projection operators
appearing in the definition of the restricted Schur polynomials.}

{\vskip 0.2cm}

\subsection{Operators dual to a Two Sphere Giant System}

There are six scalar fields $\phi^i{}_{ab}$ taking values in the
adjoint of $u(N)$ in ${\cal N}=4$ super Yang Mills theory.
Assemble these scalars into the three complex combinations
$$
  Z=\phi_1+i\phi_2 ,\qquad Y=\phi_3+i\phi_4 ,\qquad X=\phi_5+i\phi_6 \, .
$$
We will study operators built using $O(N)$ $Z$ and $O(N)$ $Y$ fields.
We always use $n$ to denote the number of $Z$s, $m$ to denote the number of $Y$s and will often refer to the $Y$ fields as ``impurities''.
These operators have a large ${\cal R}$-charge and belong to the $SU(2)$ sector of the theory.
As a consequence of this large ${\cal R}$-charge, non-planar contributions to the correlation functions of these
operators are not suppressed at large $N$.
The computation of the anomalous dimensions of these operators is then a problem of considerable complexity.
This problem has been effectively handled by new methods which employ group representation theory, allowing one to sum all
diagrams (planar and non-planar) contributing.
The new methods provide bases for the local operators which diagonalize the free two point function and which have
highly constrained mixing at the quantum level.
For the applications that we have in mind, these bases are clearly far superior to the trace basis.
Mixing between operators in the trace basis with this large ${\cal R}$-charge is completely unconstrained even at the level
of the free theory.
The particular basis we employ is provided by the restricted Schur polynomials
$$
\chi_{R,(r,s)}(Z^{\otimes \, n},Y^{\otimes \, m})
=
{1\over n!m!}\sum_{\sigma\in S_{n+m}}{\rm Tr}_{(r,s)}(\Gamma_R(\sigma))
Z^{i_1}_{i_{\sigma(1)}}\cdots Z^{i_n}_{i_{\sigma(n)}}Y^{i_{n+1}}_{i_{\sigma(n+1)}}\cdots Y^{i_{n+m}}_{i_{\sigma(n+m)}}\, .
$$
$R$ is a Young diagram with $n+m$ boxes or equivalently an
irreducible representation of $S_{n+m}$. $r$ is a Young diagram
with $n$ boxes or equivalently an irreducible representation of
$S_n$ and $s$ is a Young diagram with $m$ boxes or equivalently an
irreducible representation of $S_m$. The $S_n$ subgroup acts on
$1,2,...,n$ and therefore permutes indices belonging to the $Z$s.
The $S_m$ subgroup acts on $n+1,n+2,...,n+m$ and hence permutes
indices belonging to the $Y$s. Taken together $(r,s)$ specify an
irreducible representation of $S_n\times S_m$. ${\rm Tr}_{(r,s)}$
is an instruction to trace over the subspace carrying the
irreducible representation\footnote{In general, because $(r,s)$
can be subduced more than once, we should include a multiplicity
index. We will not write or need this index in this article.}
 $(r,s)$ of $S_n\times S_m$ inside the carrier space for irreducible representation $R$ of $S_{n+m}$.
This trace is easily realized by including a projector $P_{R\to
(r,s)}$ (from the carrier space of $R$ to the carrier space of
$(r,s)$) and tracing over all of $R$, i.e. ${\rm Tr}(P_{R\to
(r,s)}\Gamma_R(\sigma))$.

We are interested in operators dual to sphere giant gravitons that wrap a 3-sphere in the $S^5$ of the AdS$_5\times$S$^5$ background.
The two sphere giant system is described by restricted Schur polynomials labeled by Young diagrams with at most two columns.
Further, the number of $Z$s in the operator is $\alpha N$ where $0 < 2-\alpha\equiv\zeta < 1$ and the number of $Y$s is fixed to be $O(N)$.
We have chosen the number of $Z$s so that both columns of the Young diagram are long, that is, they both always have $O(N)$ boxes.
We will explain our choice for the number of $Y$s below.
It is reasonable to assume that these restricted Schur polynomials are all excitations 
(including BPS states and non-BPS states) of the two giant system.

The mixing of these operators with restricted Schur polynomials that have $n\ne 2$ columns (or of even more general shape)
is suppressed at least by a factor of order ${1\over\sqrt{N}}$\footnote{Here we are talking about mixing at the quantum level.
There is no mixing in the free theory. This suppression factor is equal to 1 over the square root of the number of boxes in
second column of the Young diagram, and it is for this reason that we must ensure that the second column in the Young diagram
has $O(N)$ boxes.}.
This factor arises from the normalization of the restricted Schur polynomials.
For example, the three column restricted Schur polynomials (with one short column - mixing is greatest for this type of operator)
have a two point function which is smaller than the two point function of the two column restricted Schur polynomials by a factor of order
${1\over N}$\cite{Koch:2010gp}.
Thus, at large $N$ the two column restricted Schur polynomials do not mix with other operators, which is a huge simplification.
This is the analog of the statement that for operators with a dimension of $O(1)$, different trace structures do not mix at large $N$.
The fact that the two column restricted Schur polynomials are a decoupled sector at large $N$ is expected: these operators correspond
to a well defined stable semi-classical object in spacetime (the two giant system).
In this large $N$ limit, and $b_0 \sim O(N)$, we can decouple the two giant system from the three giant system.
$n$ column restricted Schur polynomials are also a decoupled sector at large $N$, for the same reason.

We will end this section with a few comments on notation. The $m$
impurity operators are built using $n$ $Z$s and $m$ $Y$s. There
are three Young diagrams labeling the restricted Schur polynomial,
$\chi_{R,(r,s)}$. The representation $r$ (which specifies an
irreducible representation of $S_n$) is specified by stating the
number of rows with two boxes ($=b_0$) and the number of rows with
a single box ($=b_1$). The representation $s$ specifies an
irreducible representation of $S_m$ that has at most two columns,
while $R$, which specifies an irreducible representation of
$S_{n+m}$, can be specified in terms of $r$ by stating which boxes
in $R$ are to be removed to obtain $r$. Once $R$ has been given,
there are a finite number of possible labels $R,(r,s)$ even in the
limit $N\to\infty$. To illustrate the problem, it is useful to
first study examples with small $m$ of order $O(1)$. In the next
subsection we will be interested in the case $m=4$. In this case
there are nine operators that can be produced once $R$ is given.
These operators are given below
$$
\chi_A(b_0,b_1)=\chi_{\tiny \yng(2,2,2,2,2,2,1,1,1,1,1)\, \yng(2,2,2,2,2,2,1)\, \yng(1,1,1,1)}(Z,Y)
\qquad
\chi_B(b_0,b_1)=\chi_{\tiny \yng(2,2,2,2,2,2,1,1,1,1,1)\, \yng(2,2,2,2,2,1,1,1)\, \yng(1,1,1,1)}(Z,Y)
$$
$$
\chi_C(b_0,b_1)=\chi_{\tiny \yng(2,2,2,2,2,2,1,1,1,1,1)\,\yng(2,2,2,2,2,1,1,1)\, \yng(2,1,1)}(Z,Y)
\qquad
\chi_D(b_0,b_1)=\chi_{\tiny \yng(2,2,2,2,2,2,1,1,1,1,1)\, \yng(2,2,2,2,1,1,1,1,1)\, \yng(1,1,1,1)}(Z,Y)
$$
$$
\chi_E(b_0,b_1)=\chi_{\tiny \yng(2,2,2,2,2,2,1,1,1,1,1)\, \yng(2,2,2,2,1,1,1,1,1)\, \yng(2,1,1)}(Z,Y)
\qquad
\chi_F(b_0,b_1)=\chi_{\tiny \yng(2,2,2,2,2,2,1,1,1,1,1)\, \yng(2,2,2,2,1,1,1,1,1)\, \yng(2,2)}(Z,Y)
$$
$$
\chi_G(b_0,b_1)=\chi_{\tiny \yng(2,2,2,2,2,2,1,1,1,1,1)\, \yng(2,2,2,1,1,1,1,1,1,1)\, \yng(1,1,1,1)}(Z,Y)
\qquad
\chi_H(b_0,b_1)=\chi_{\tiny \yng(2,2,2,2,2,2,1,1,1,1,1)\, \yng(2,2,2,1,1,1,1,1,1,1)\, \yng(2,1,1)}(Z,Y)
$$
$$
\chi_I(b_0,b_1)=\chi_{\tiny \yng(2,2,2,2,2,2,1,1,1,1,1)\, \yng(2,2,1,1,1,1,1,1,1,1,1)\, \yng(1,1,1,1)}(Z,Y)
$$
For the operators we study the number of rows with two boxes in
the first and second Young diagram labels scale as $N$. Thus, the
length of the columns in the first two labels are taken to be very
large at large $N$. The third label is exactly as shown for the
$m=4$ case even at large $N$. The number of boxes removed from
each column of the first label to obtain the second label are
exactly as shown, even at large $N$. In what follows
we will study the case for which $m$ is of order
$O(N)$. Thus, in general, the column lengths of all the three
Young diagrams are of order $O(N)$ so that they are very long
and thus are not shown in the figure.

\subsection{Simplified Action of the Dilatation Operator}

The action of the one loop dilatation operator in the $SU(2)$ sector\cite{Beisert:2003tq}
$$
D = - g_{\rm YM}^2 {\rm Tr}\,\big[ Y,Z\big]\big[ \partial_Y ,\partial_Z\big]
$$
on the restricted Schur polynomial has been studied in \cite{Koch:2010gp}.
We will find it convenient to work with operators normalized to give a unit two point function.
Towards this end, note that the two point function for restricted Schur polynomials has been computed in \cite{Bhattacharyya:2008rb}
$$
\langle\chi_{R,(r,s)}(Z,Y)\chi_{T,(t,u)}(Z,Y)^\dagger\rangle =
\delta_{R,(r,s)\,T,(t,u)}f_R {{\rm hooks}_R\over {\rm hooks}_{r}\, {\rm hooks}_s}\, .
$$
In this expression $f_R$ is the product of the weights in Young diagram $R$ and ${\rm hooks}_R$ is the product of the hook
lengths of Young diagram $R$.
Using this result, the normalized operators $O_{R,(r,s)}(Z,Y)$ can be obtained from
$$
\chi_{R,(r,s)}(Z,Y)=\sqrt{f_R \, {\rm hooks}_R\over {\rm hooks}_r\, {\rm hooks}_s}O_{R,(r,s)}(Z,Y)\, .
$$
An important intermediate result is
{\small
\begin{eqnarray}
\label{basicreslt}
D \, \chi_{R,(r,s)}(Z^{\otimes \, n},Y^{\otimes \, m})=&&
{g_{\rm YM}^2\over (n-1)!(m-1)!}\sum_{\psi\in S_{n+m}}
\Tr_{(r,s)}\left(\Gamma_R ((n,n+1) \psi -\psi (n,n+1))\right)\times\cr
\times Z^{i_1}_{i_{\psi (1)}}&&
\cdots
Z^{i_{n-1}}_{i_{\psi (n-1)}}
(YZ-ZY)_{i_{\psi (n)}}^{i_n}\delta^{i_{n+1}}_{i_{\psi (n+1)}}
Y^{i_{n+2}}_{i_{\psi (n+2)}}\cdots Y^{i_{n+m}}_{i_{\psi (n+m)}}
\, .
\end{eqnarray}
}
Notice that one of the indices associated to a $Z$ field and one of the
indices associated to a $Y$ field participates. This is reflected in the formula
below by the appearance of the group element $\Gamma_R(n,n+1)$ which does not belong
to the $S_m\times S_n$ subgroup.
In terms of these normalized operators
$$
DO_{R,(r,s)}(Z,Y)=\sum_{T,(t,u)} N_{R,(r,s);T,(t,u)}O_{T,(t,u)}(Z,Y)
$$
\begin{eqnarray}
\label{dilat}
N_{R,(r,s);T,(t,u)}&&= - g_{YM}^2\sum_{R'}{c_{RR'} d_T n m\over d_{R'} d_t d_u (n+m)}
\sqrt{f_T \, {\rm hooks}_T\, {\rm hooks}_r \, {\rm hooks}_s \over f_R \, {\rm hooks}_R\, {\rm hooks}_t\, {\rm hooks}_u}\times
\\
\nonumber
&&\times\Tr\Big(\Big[ \Gamma_R((n,n+1)),P_{R\to (r,s)}\Big]I_{R'\, T'}\Big[\Gamma_T((n,n+1)),P_{T\to (t,u)}\Big]I_{T'\, R'}\Big) \, .
\end{eqnarray}
The $c_{RR'}$ is the weight of the corner box removed from Young
diagram $R$ to obtain diagram $R'$, and similarly $T'$ is a Young
diagram obtained from $T$ by removing a box.
The interwiner $I_{AB}$ is a map from the carrier space of irreducible representation $A$ to
the carrier space of irreducibe representation $B$. Consequently, $A$ and $B$ must be Young diagrams
of the same shape. The intertwiner operator relevant for our study will be evaluated in sections
\ref{compute_A}, \ref{young_S}.

This last expression
has been evaluated in \cite{Koch:2010gp} for the case of two
impurities and in \cite{VinceKate} for the case of three or four
impurities. The results are rather complicated.
However, in the limit that $N-b_0=O(N)$, $b_0=O(N)$ and $b_1=O(N)$
the dynamics simplifies considerably. For $m=4$ impurities, the
action of the dilatation operator becomes
\begin{equation}
DO_A(b_0,b_1)=g_{YM}^2\Delta O_A(b_0,b_1)\times O\left({1\over b_1}\right)
\end{equation}
\begin{equation}
DO_B(b_0,b_1) =-{3\over 2}g_{YM}^2\Delta O_B(b_0,b_1) +{\sqrt {3}\over 2}g_{YM}^2\Delta O_C(b_0,b_1)
\end{equation}
\begin{equation}
DO_C(b_0,b_1) = {\sqrt {3}\over 2}g_{YM}^2\Delta O_B(b_0,b_1)-{1\over 2}g_{YM}^2\Delta O_C(b_0,b_1)
\end{equation}
\begin{equation}
DO_D(b_0,b_1) =-2g_{YM}^2\Delta O_D(b_0,b_1)+{2\over \sqrt{3}}g_{YM}^2\Delta O_E(b_0,b_1)
\end{equation}
\begin{equation}
DO_E(b_0,b_1) =-2g_{YM}^2\Delta O_E(b_0,b_1)+{2\over \sqrt{3}}\Delta O_D(b_0,b_1)+{2\sqrt{6}\over 3}\Delta O_F(b_0,b_1)
\end{equation}
\begin{equation}
DO_F(b_0,b_1) =-2\Delta O_F(b_0,b_1) +{2\sqrt{6}\over 3}\Delta O_E(b_0,b_1)
\end{equation}
\begin{equation}
DO_G(b_0,b_1) =-{3\over 2}g_{YM}^2\Delta O_G(b_0,b_1)+{\sqrt{3}\over 2}g_{YM}^2\Delta O_H(b_0,b_1)
\end{equation}
\begin{equation}
DO_H(b_0,b_1) =-{1\over 2}g_{YM}^2\Delta O_H(b_0,b_1) +{\sqrt{3}\over 2}g_{YM}^2\Delta O_G(b_0,b_1)
\end{equation}
\begin{equation}
DO_I(b_0,b_1)=g_{YM}^2\Delta O_I(b_0,b_1)\times O\left({1\over b_1}\right)
\end{equation}
where
\begin{eqnarray}
\nonumber
\Delta O_X(b_0,b_1)&&=
\sqrt{(N-b_0-b_1)(N-b_0)}(O_X(b_0+1,b_1-2)+O_X(b_0-1,b_1+2))\\
&&-(2N-2b_0-b_1) O_X (b_0,b_1)\, .
\end{eqnarray}
with $X=A,B,...,I$. A slightly different limit was considered in
\cite{VinceKate} - there it was assumed that $b_1\sim
O(\sqrt{N})$. Since $b_1$ sets the difference in angular momentum
of the two giants and the angular momentum of the two giants sets
their radii \cite{Myers:1999ps}, this limit set the distance
between the two giants to be string size. This is the natural
limit to look for (open) stringy excitations of the giants. As we
will see, the limit we consider here will allow us to reproduce
spectra obtained by numerically diagonalizing the dilatation
operator. We can easily identify combinations of operators that
are annihilated by $D$. Apart from $O_A(b_0,b_1)$ and
$O_I(b_0,b_1)$ we have $O_B(b_0,b_1)+\sqrt{3}O_C(b_0,b_1)$,
$O_D(b_0,b_1)+\sqrt{3}O_E(b_0,b_1)+\sqrt{2}O_F(b_0,b_1)$ and
$O_G(b_0,b_1)+\sqrt{3}O_H(b_0,b_1)$. If we set
$\sqrt{3}O_B(b_0,b_1)-O_C(b_0,b_1)\equiv O_{B-C}(b_0,b_1)$,
$\sqrt{2}O_D(b_0,b_1)-O_F(b_0,b_1)\equiv O_{D-F}(b_0,b_1)$,
$O_D(b_0,b_1)-\sqrt{3}O_E(b_0,b_1)+\sqrt{2}O_F(b_0,b_1)\equiv
O_{DF-E}(b_0,b_1)$ and $\sqrt{3}O_G(b_0,b_1)-O_H(b_0,b_1)\equiv
O_{G-H}(b_0,b_1)$, we have
{\small
\begin{eqnarray}
DO_{B-C}(b_0,b_1) = &&-2g_{YM}^2 \Delta O_{B-C}(b_0,b_1)\cr
DO_{D-F}(b_0,b_1) = &&-2g_{YM}^2 \Delta O_{D-F}(b_0,b_1)\cr
DO_{DF-E}(b_0,b_1) = &&-4g_{YM}^2\Delta O_{DF-E}(b_0,b_1)\cr
DO_{G-H}(b_0,b_1) = &&-2g_{YM}^2 \Delta O_{G-H}(b_0,b_1)
\end{eqnarray}
}
An exact diagonalization of the dilatation operator gives the spectrum of nine harmonic oscillators.
Five of the oscillators have a level spacing $\omega =0$, three have a level spacing $\omega =8g_{YM}^2$
and one has a level spacing $\omega =16g_{YM}^2$.
These results can all be obtained from the simplified action of the dilatation operator given above.
The generic form of the equation coming from the action of the dilatation operator is
\begin{eqnarray}
DO(b_0,b_1) =&& -\alpha g_{YM}^2[\sqrt{(N-b_{0})(N-b_{0}-b_{1})}(O (b_{0}+1,b_{1}-2)+O (b_{0}-1,b_{1}+2))\nonumber \\
 &&-(2N-2b_{0}-b_{1})O (b_{0},b_{1})]\label{oscillator02}
\end{eqnarray}
which corresponds to an oscillator of level spacing $4\alpha g_{YM}^2$. The
derivation of this oscillator level spacing is in Appendix
\ref{Appendix_osc}. The goal of this article is to derive these
simplified equations directly in the $b_1\sim O(N)$ limit
and then to diagonalize them analytically. The main difficulty in
deriving these equations is in the explicit construction of the
projectors used in defining the restricted Schur polynomials.

\subsection{Young's Orthogonal Representation}

There is a representation for the symmetric group, Young's
orthogonal representation, that we will make extensive use of. To
define the representation, we will give the rule which determines
matrix elements of the matrices representing adjacent
permutations, that is, permutations of the form $(i,i+1)$, where
$i$ denotes the $i$th box. The matrix representing any other
element of the group can easily be constructed as some product of
the matrices representing the adjacent permutations. We will use
the Young-Yamonouchi basis in which the boxes in the Young diagram
are numbered. We choose our conventions so that if the boxes are
removed according to the above numbering, removing box 1 first,
box 2 second and so on, at each step one must always obtain a
legal Young diagram. Thus, for the two Young diagrams shown below,
the diagram on the right corresponds to a valid state; the one on
the left does not
$$
\young(14,32,5)\qquad \young(54,32,1)\, .
$$
The dimension of the $S_5$ irreducible representation labeled by $\yng(2,2,1)$ is 5. The reader can check that there are five possible
labels each giving a state in the carrier space of this irreducible representation. Recall that to each labeled box we can associate
a weight. Denote the weight of the box labeled $i$ by $c_i$. If box $i$ is in row $b$ and column $a$, it has $c_i=N+a-b$.
We can now state the rule for the action of adjacent transpositions: If $(i,i+1)$ acts on a given state, it gives the
same state back with coefficient ${1\over c_i-c_{i+1}}$ and it gives the state corresponding to the Young diagram with $i$ and $i+1$
swapped with coefficient\footnote{Here to define the representation of
$S_n$ we have used the weights $c_i$ which depend on $N$. The $S_n$ group should know nothing about $N$. Since only differences between
weights appear, this is indeed the case.} $\sqrt{1-{1\over (c_i-c_{i+1})^2}}$. Here are a few examples
$$
\Gamma_{\yng(2,2,1)}\left( (12)\right)|\young(53,42,1)\rangle =
 -{1\over 2}|\young(53,42,1)\rangle
 +{\sqrt{3}\over 2}|\young(53,41,2)\rangle
$$
$$
\Gamma_{\yng(2,2,1)}\left( (12)\right)|\young(53,41,2)\rangle =
 {1\over 2}|\young(53,41,2)\rangle
 +{\sqrt{3}\over 2}|\young(53,42,1)\rangle
$$
$$
\Gamma_{\yng(2,2,1)}\left( (12)\right)|\young(52,41,3)\rangle = -|\young(52,41,3)\rangle
$$
Young's orthogonal representation is particularly useful because
it simplifies dramatically when\footnote{Recall that $m$ is the
number of $Y$s.} ${m\over b_1}\ll 1$. Since the generic operator
we consider has $b_1=O(N)$, we will consider $m=\gamma N$
impurities with $\gamma \sim O(N^0)\ll 1$. In this simplification,
we can consider the ratio $\alpha/\gamma \sim O(N^0) \gg 1$ as a
fixed large number, which does not need to scale as $N$ to some
positive power. Indeed, if the boxes $i$ and $i+1$ are in the same
column, $i+1$ must sit above $i$ so that
\begin{eqnarray}
\Gamma_R \left( (i,i+1)\right)|{\rm same\, column\, state}\rangle
= -|{\rm same\, column\, state}\rangle
\end{eqnarray}
If $i$ and $i+1$ are in different columns, then since $b_1=O(N)$, even if we stack all of the impurities
in the longer column $c_i-c_{i+1}$ must itself be $O(N)$. In this case, at large $N$ replace ${1\over c_i-c_{i+1}}=O(b_1^{-1})$ by 0 and
$\sqrt{1-{1\over (c_i-c_{i+1})^2}}=1-O(b_1^{-1})$ by 1 so that
\begin{eqnarray}
\Gamma_R \left( (i,i+1)\right)|{\rm different\, column\,
state}\rangle = |{\rm swapped\, different\, column\, state}\rangle
\end{eqnarray}
The notation in this last equation is indicating two things: $i$ and $i+1$ are in different columns and the states on the two
sides of the equation differ by swapping the $i$ and $i+1$ labels.
All of the states which enter the trace in the restricted Schur polynomial (and hence in the dilatation operator) belong to
a particular $S_n\times S_m$ subspace. As we explain now, it is possible to directly extract part of this subspace and hence one need
not work with the general state in $R$. Start numbering our states using a Young-Yamonouchi basis. We can obtain $r$ by removing boxes
from $R$. If we only number the first $m$ boxes and further only boxes that are to be removed are numbered, then each partially
labeled Young diagram stands for a collection of states, all belonging to the correct $S_n$ subspace. Thus, all we need to do now
is to take the correct combinations of these states (i.e. of the partially labeled Young diagrams) to get the required $S_m$ subspace.
The group element $\Gamma(n,n+1)$ appears in the expression of the dilatation operator, and it acts on one of the $m$ boxes associated
to the impurities and one box associated to the $Z$s. Taken together, these facts imply that
all of the operators that we need to consider only have a non-trivial action on the first $m+1$ boxes. Thus we will not label
all of the boxes – it is good enough to label the first $m+1$ boxes.
Bear in mind that each partially labeled Young diagram corresponds to a collection of states.
An example illustrating these rules: (assuming just $m=2$ impurities so that we only label 3 boxes)
{\small
$$
\Gamma_R \left(
(1,2)\right)\left.\Big|\young(\,\,,\,\,,\,\,,\,\,,\,\,,\,1,\,,\,,\,,\,,\,,3,2)\right\rangle
=\left.\Big|\young(\,\,,\,\,,\,\,,\,\,,\,\,,\,2,\,,\,,\,,\,,\,,3,1)\right\rangle
\qquad\qquad \Gamma_R \left(
(1,2)\right)\left.\Big|\young(\,\,,\,\,,\,\,,\,\,,\,\,,\,\,,\,,\,,\,,\,,3,2,1)\right\rangle
=-\left.\Big|\young(\,\,,\,\,,\,\,,\,\,,\,\,,\,\,,\,,\,,\,,\,,3,2,1)\right\rangle
$$
}
Thus, the representations of the symmetric group simplify
dramatically in this limit.

\subsection{Computation of the Dilatation Operator}
\label{compute_A}

The dilatation operator includes the coefficient
$$
-g_{YM}^2{c_{RR'}d_T d_{R^{m+1}}nm\over d_{R'}d_t d_u (n+m)}
\sqrt{f_T {\rm hooks}_T{\rm hooks}_r{\rm hooks}_s\over f_R {\rm hooks}_R{\rm hooks}_t{\rm hooks}_u}\, .
$$
In the above formula $d_{R^{m+1}}$ labels an irreducible representation of $S_{n-1}$ and is obtained from $R$
(which labels an irreducible representation of $S_{n+m}$) by removing $m+1$ boxes.
The dimension factor $d_{R^{m+1}}$ is included in the above coefficient for convenience - it naturally appears when we evaluate the trace
\begin{equation}
\Tr\Big(\Big[ \Gamma_R((n,n+1)),P_{R\to (r,s)}\Big]I_{R'\, T'}\Big[\Gamma_T((n,n+1)),P_{T\to (t,u)}\Big]I_{T'\, R'}\Big) \, .
\label{trace}
\end{equation}
There are different choices for the specific $R^{m+1}$s but, in the limit that we consider, they all give the same contribution.
In the limit we consider
$$
{{\rm hooks}_T{\rm hooks}_r\over {\rm hooks}_R{\rm hooks}_t}=1 \qquad {d_T\over d_t}=2^m
$$
$$
{n\over n+m}={\alpha \over
{\alpha+\gamma}}\to 1 \qquad {d_{R^{m+1}}\over d_{R'}}=2^{-m}
\qquad d_u = {m!\over {\rm hooks}_u}
$$
so that
\begin{equation}
-g_{YM}^2{c_{RR'}d_T d_{R^{m+1}}nm\over d_{R'}d_t d_u (n+m)}
\sqrt{f_T {\rm hooks}_T{\rm hooks}_r{\rm hooks}_s\over f_R {\rm
hooks}_R{\rm hooks}_t{\rm hooks}_u} =-{g_{YM}^2\sqrt{c_{RR'}c_{TT'}}\sqrt{{\rm
hooks}_s{\rm hooks}_u}\over (m-1)!}\, .
\end{equation}
All that remains in the evaluation of $D$ is the computation of the trace (\ref{trace}).
Recall that, because we have only labeled $m+1$ boxes, when we talk about a ``state'' we actually mean a collection of states.
All of the operators we specify below have the same action on the collection of states being considered.
Since we are talking about collections of states the reader should bear in mind that we have often suppressed a factor
of the identity matrix acting on the collection of states.
We do this in all formulas that follow.
It is straight forward to verify that for the case of $m$ impurities, the intertwiners are $2^{m+1}\times 2^{m+1}$ matrices, that there
are four possible intertwiners and that their non zero elements are\footnote{This next formula
nicely illustrates our conventions. The indices
$i$ and $j$ do not select a unique state - they select a collection of states. Thus, there is an identity matrix acting on the collection
of states which is suppressed on the right hand side of this equation.}
$$
(I_1)_{ij}=1,\qquad i=j=1,2,3,...,2^m
$$
(this intertwiner is used when evaluating terms in $N_{R,(r,s);T,(t,u)}$ for which $R=T$ and the box is removed from the first column
of $R$ to obtain $R'$)
$$
(I_2)_{ij}=1,\qquad i=j=2^m+1,2^m+2,2^m+3,...,2^{m+1}
$$
(again $R=T$ but now the box is removed from the second column of $R$ to obtain $R'$) for the terms with no changes in $R$ and
$$
(I_3)_{ij}=1,\qquad j=i+2^m=2^m+1,2^m+2,2^m+3,...,2^{m+1}
$$
(now $R\ne T$ and the box is removed from the first column in $R$ to obtain $R'$ and from the second column in $T$ to obtain $T'$)
$$
I_4=(I_3)^T
$$
(again $R\ne T$ and the box is removed from first column in $T$ to obtain $T'$ and from the second column in $R$ to obtain $R'$).
The non zero elements of $\Gamma \left((n,n+1)\right)$ are
$$
 \Gamma \left((n,n+1)\right)_{ij}=-1,\qquad i=j=1+2p \qquad p=0,1,...,2^{m-1}-1
$$
$$
 \Gamma \left((n,n+1)\right)_{ij}=-1,\qquad i=j=2^m+2+2p \qquad p=0,1,...,2^{m-1}-1
$$
$$
 \Gamma \left((n,n+1)\right)_{ij}=(-1)^{m-1},\qquad i=2+2p,\quad j=2^m+1+2p \qquad p=0,1,2^m-1
$$
$$
 \Gamma \left((n,n+1)\right)_{ij}=(-1)^{m-1},\qquad j=2+2p,\quad i=2^m+1+2p \qquad p=0,1,2^m-1
$$
Next one needs to compute the projector and then the trace. Without a few new ideas it is not possible
to find formulas for general $m$ for the projectors used in constructing restricted Schur
polynomials labeled by Young diagrams that have two long columns. These new ideas are most easily
developed by studying restricted Schur polynomials labeled by Young diagrams
with two long rows. This is the case that we study next.

\section{Two AdS Giant Gravitons}

{    In this section we first review the action of the dilatation
     operator on a system of two AdS giant gravitons. This two giant graviton system is described by restricted Schur polynomials labeled by Young diagrams
     that have two long rows. We again argue that the action of the dilatation operator simplifies considerably in a particular limit.
     The general construction of the dilatation operator directly in this limit, is considered.
     We argue that Young's orthogonal representation for the symmetric group in this setting reduces to a representation
     that is known and has been studied. Using these results we introduce an operator $d^\dagger d$ whose eigenspaces
     are irreducible representations of the symmetric group. Thus, a general explicit construction of the projection
     operators appearing in the definition of the restricted Schur polynomials is reduced to solving the eigenvalue problem
     of $d^\dagger d$.
     We will use these results to give an explicit formula for the dilatation operator acting on a general system of two AdS
     giant gravitons in the next section.}

{\vskip 0.2cm}

\subsection{Operators dual to a Two AdS Giant System}

We have focused on operators dual to sphere giant gravitons that
wrap a 3-sphere in the $S^5$ of the AdS$_5\times$S$^5$ background.
In this section we would like to focus on giant gravitons that are
large in the AdS directions but continue to carry angular momentum
on the $S^5$. The two AdS giant system is described by restricted
Schur polynomials labeled by Young diagrams with at most two rows.
We will continue to assume that the number of $Z$s in the operator
is $\alpha N$ where $\alpha\sim O(N^0)>1$ and the number of $Y$s
is fixed to be $\gamma N$ with $\gamma \sim O(N^0)\ll 1$. The
mixing of these operators with restricted Schur polynomials that
have $n\ne 2$ rows (or of even more general shape) is again
suppressed at least by a factor of order
${1\over\sqrt{N}}$\footnote{Here we are again talking about mixing
at the quantum level. There is again no mixing in the free
theory.}. The fact that the two row restricted Schur polynomials
are a decoupled sector at large $N$ is again expected: these
operators correspond to a well defined stable semi-classical
object in spacetime (the two AdS giant system).

We only need to make very minor changes in our notation. There are still three Young diagrams labeling the restricted Schur polynomial,
$\chi_{R,(r,s)}$.
The representation $r$ (which specifies an irreducible representation of $S_n$) is specified by stating the number of columns
with two boxes ($=b_0$) and the number of columns with a single box ($=b_1$).
The representation $s$ specifies an irreducible representation of $S_m$ that has at most two rows, while
$R$, which specifies an irreducible representation of $S_{n+m}$, can be specified in terms of $r$ by stating which boxes in $R$ are to be removed to
obtain $r$. Recall that in the sphere giant system,
once $r$ has been given there are a finite number of possible labels $R,(r,s)$ even in the limit $N\to\infty$.
This is also the case for the AdS giant system.

The two AdS giant system and two sphere giant system have an
important difference. For the two sphere giant system the
parameter $b_1$ was bounded from above and the parameter $b_0$ was
bounded from below. Indeed, the largest possible value for $b_1$
and the smallest possible value for $b_0$ were obtained when the
first column contains $N$ boxes. Of course these bounds on
$b_0,b_1$ are a consequence of the fact that the number of boxes
in any given column is bounded by $N$. For the AdS giant system
the number of boxes in the first row is not bounded, so that $b_0$
can range all the way down to 0. For the sphere giant case the
fact that $b_0$ was bounded from below, and the fact that this
bound was $O(N)$, implied a clean decoupling of the two sphere
giant system. For the AdS giants, when $b_0\ll N$ we are
transitioning to the state of one giant graviton plus strings, or
plus Kaluza-Klein gravitons, or both. The two AdS giant system is
still effectively decoupled because it would take a very long
time\footnote{What we mean by this is that we need to apply the
Hamiltonian $O(N)$ times before we leave the two giant sector.}
before we decay from the two giant sector. This point does however
need to be considered when we study the dilatation operator
numerically. In our numerical studies we will only consider Young
diagrams $R$ with two rows. This is strictly speaking not always
justified because when the second row contains $O(1)$ boxes mixing
with Young diagrams with one long row and more than one short row
is not suppressed. We will discuss this point further below (see
section 7) and argue that it does not affect the accuracy of our
results.

\subsection{Dilatation Operator}

For the two sphere giant system we have seen that when $b_1\gg 1$ the action of the dilatation operator simplifies
dramatically. Is this also the case for the action of the
dilatation operator on the two AdS giant system? To compute the
action of the dilatation operator we need to again evaluate
(\ref{dilat}), but now for the case that $R$ and $T$ have two
rows. The result of this computation for the case of two
impurities is given in Appendix \ref{A}. There are a number of
points worth noting. First, there is a simple relation between the
sphere giant and AdS giant results. To obtain the action of the
dilatation operator on the AdS giant system from the action of the
dilatation operator on the sphere giant system one simply replaces
the sphere weights $N-a$ to $N+a$. This is completely explicit if
the reader compares the first two expressions in Appendix \ref{A}
to the expressions appearing in \cite{Koch:2010gp}. In the last
two expressions of Appendix \ref{A}, this is not manifest because
we have combined terms. Second, the numerical spectrum for the AdS
giant system is surprisingly similar to the spectrum obtained for
the sphere giant system\cite{Koch:2010gp}. 3/4 of the states are
massless while 1/4 of the states match those of an oscillator with
an energy spacing of 8$g_{YM}^2$. This looks identical to what was
obtained for the sphere giants. One important difference however,
is the fact that for the sphere giant system the levels were of
the form $8g_{YM}^2 n$ with $n=1,2,3,...$ where as for the AdS
giant system the levels are of the form $4g_{YM}^2 + 8g_{YM}^2 n$
with $n=1,2,3,...$. Particularly for the low levels, it seems that
these results can be trusted, because the operator with good scale
dimension receives no contribution\footnote{The contribution is
zero to the accuracy of our numerical diagonalization.} from
restricted Schur polynomials with one very long column and one
short column. Finally, perhaps the most important result we find
is that in the limit that ${b_1\over m}\gg 1$ the action of the dilatation operator
reduces to a collection of equations of the form
\begin{eqnarray}
DO(b_0,b_1) =&& -\alpha g_{YM}^2[\sqrt{(N+b_{0})(N+b_{0}+b_{1})}(O (b_{0}+1,b_{1}-2)+O (b_{0}-1,b_{1}+2))\nonumber \\
 &&-(2N+2b_{0}+b_{1})O (b_{0},b_{1})]\, .\label{oscillator01}
\end{eqnarray}
This corresponds to an oscillator of level spacing $4\alpha g_{YM}^2$. More
detailed derivations are in Appendix \ref{Appendix_osc}. This result
is again related to our previous result for the sphere giant system, by
replacing for example $N-b_0\to N+b_0$.

\subsection{Young's Orthogonal Representation}
\label{young_S}

In the case of AdS giants, Young's orthogonal representation
reduces to a representation which has already been studied in the
mathematics literature \cite{GelPair}. The reader may wish to
consult Appendix \ref{rep} where we review a little of the
relevant background. The idea is to define a map from a labeled
Young diagram to a monomial. We will consider the case of $m$ $Y$
fields. In this case, consider a labeled Young diagram that has
$m$ boxes labeled\footnote{Up to now we have labeled $m+1$ boxes
when considering $m$ impurities. For the present discussion where
we want to understand how to decompose into irreducible $S_m$
representations it is more convenient to label only $m$ boxes.};
the labels are distributed arbitrarily between the upper and lower
rows. Ignore the boxes that appear in the lower row. For boxes
labeled $i$ in the upper row include a factor of $x_i$ in the
monomial. If none of the boxes in the first row are labeled, the
Young diagram maps to 1. Thus, for example, when $m=4$
{\small
$$
{\small \young({\,}{\,}{\,}{\,}{\,}{\,}{\,}{\,}{\,}{3},{\,}{\,}{4}{2}{1})}\leftrightarrow x_3
\qquad
{\small \young({\,}{\,}{\,}{\,}{\,}{\,}{\,}{3}{2}{1},{\,}{\,}{\,}{\,}{4})}\leftrightarrow
x_1 x_2 x_3
$$
}
The symmetric group acts by permuting the labels on the factors in the monomial. Thus, for example,
$ (12)x_1 x_3 = x_2 x_3 $. This defines a reducible representation of the symmetric group, $S_m$. It is clear that
the operator
\begin{equation}
d=\sum_{i=1}^m {\partial\over \partial x_i}
\end{equation}
commutes with the action of the symmetric group. This operator was
introduced by Dunkl in the study of intertwining functions
\cite{Dunkl}. It acts on the monomials by producing the sum of
terms that can be produced by dropping one factor at a time. For
example
$$
d(x_1 x_2 x_3) = x_2 x_3 + x_1 x_3 + x_1 x_2 \, .
$$
The adjoint\footnote{Consult Appendix \ref{rep} for details on the
inner product on the space of monomials.} $d^\dagger$ produces the
sum of monomials that can be obtained by appending a factor,
without repeating any of the $x_i$s (this is written for $m=4$
impurities but the generalization to any $m$ is obvious)
$$
  d^\dagger (x_1 x_2)= x_1 x_2 x_3 + x_1 x_2 x_4\, .
$$
The fact that $d$ commutes with all elements of the symmetric group, implies that $d^\dagger$ will too. Indeed, take the dagger of
$$ \big[ d,\Gamma(\sigma )\big]=0 $$
to obtain (use the fact that we are working in an orthogonal representation in the next line)
$$ \big[ d^\dagger ,\Gamma(\sigma )\big]=-(\big[ d,\Gamma(\sigma^{-1} )\big])^\dagger = 0 \, .$$
Thus, $d^\dagger d$ will also commute with all the elements of the symmetric
group and consequently its eigenspaces will furnish representations
of the symmetric group. These eigenspaces are irreducible representations - consult \cite{GelPair} for further details and results.
This last fact implies that the problem of computing the projectors needed to define the restricted Schur polynomials can be replaced by
the problem of constructing projectors onto the eigenspaces of $d^\dagger d$. This amounts to solving for the eigenvectors and eigenvalues
of $d^\dagger d$. In the next section we will argue that this is a surprisingly simple problem.

For the remainder of this section we switch back to our previous convention and again label $m+1$ boxes. See section
2.4 for a careful description of our notation.
The intertwiners are again $2^{m+1}\times 2^{m+1}$ matrices, and again there are four possible
intertwiners. Their non zero elements are
$$
(I_1)_{ij}=1,\qquad i=j=1,2,3,...,2^m
$$
($R=T$ and the box is removed from first column)
$$
(I_2)_{ij}=1,\qquad i=j=2^m+1,2^m+2,2^m+3,...,2^{m+1}
$$
($R=T$ and the box is removed from second column) for the terms with no changes in $R$ and
$$
(I_3)_{ij}=1,\qquad j=i+2^m=2^m+1,2^m+2,2^m+3,...,2^{m+1}
$$
(the box is removed from first column in $R$ and from the second column in $T$)
$$
I_4=(I_3)^T
$$
(the box is removed from first column in $T$ and from second column in $R$).
The non zero elements of $\Gamma \left((n,n+1)\right)$ are
$$
 \Gamma \left((n,n+1)\right)_{ij}=1,\qquad i=j=1+2p  \qquad p=0,1,...,2^{m-1}-1
$$
$$
 \Gamma \left((n,n+1)\right)_{ij}=1,\qquad i=j=2^m+2+2p  \qquad p=0,1,...,2^{m-1}-1
$$
$$
 \Gamma \left((n,n+1)\right)_{ij}=1,\qquad i=2+2p,\quad j=2^m+1+2p \qquad p=0,1,2^m-1
$$
$$
 \Gamma \left((n,n+1)\right)_{ij}=1,\qquad j=2+2p,\quad i=2^m+1+2p \qquad p=0,1,2^m-1
$$

\subsection{Computation of the Dilatation Operator}

To get to the AdS giant case, we have ``flipped'' the Young
diagrams swapping rows and columns, and mapped the weights $N-b
\to N+b$. It is easy to check that the dimension of any Young
diagram $R$ is the same as the dimension of the flipped diagram,
and (this is the same fact that) the product of hooks is
unchanged. The weights $c_{RR'}$ are now for example $N+b_0$
instead of $N-b_0$. 

Thus, the coefficient needed to compute the dilatation operator is
\begin{equation}
-g_{YM}^2{c_{RR'}d_T d_{R^{m+1}}nm\over d_{R'}d_t d_u (n+m)}
\sqrt{f_T {\rm hooks}_T{\rm hooks}_r{\rm hooks}_s\over f_R {\rm
hooks}_R{\rm hooks}_t{\rm hooks}_u} =-{g_{YM}^2\sqrt{c_{RR'}c_{TT'}}\sqrt{{\rm
hooks}_s{\rm hooks}_u}\over (m-1)!}
\end{equation}
$d_{R^{m+1}}$ again labels an irreducible representation of $S_{n-1}$ and 
is again obtained from $R$ (which labels an irreducible representation of 
$S_{n+m}$) by removing $m+1$ boxes. The dimension factor $d_{R^{m+1}}$ is 
again included in the above coefficient for convenience - it naturally 
appears when we evaluate the trace (\ref{trace}).
All that we need to evaluate now are the traces (\ref{trace}). We
will show how to evaluate these for general $m$ in the next
section.

\section{Construction of AdS Giant Projectors and the Dilatation Operator}
\label{construction_A}

{ In this section we define a map from labeled Young diagrams to
spin chain states. The operator $d^\dagger d$ takes
     a particularly simple form and its eigenvalue value problem is solved explicitly. These results then allow a general
     construction of the projection operators used to define the restricted Schur polynomials and then ultimately of the
     dilatation operator itself. Our final formula for the dilatation operator is given as the last equation of this section.}

{\vskip 0.2cm}

We can map the labeled Young diagrams into states of a spin chain,
for general $m$. The spin at site $i$ can be in state spin up (1)
or state spin down (0). Initially in this section it is again
convenient to label only $m$ boxes. In this case, the spin chain
has $m$ sites and the box $i$ tells us the state of site $i$. If
box $i$ appears in the first row, site $i$ is in state 1; if it
appears in the second row site $i$ is in state 0. For example,
{\small
$$
\young({\,}{\,}{\,}{\,}{\,}{\,}{\,}{\,}{2}{1},{\,}{\,}{\,}{4}{3})\leftrightarrow |1100\rangle
$$
}
$d^\dagger d$ has a very simple action on this spin chain as we now explain: Introduce the states
$$
|0\rangle =\left[ \begin {array}{c} 1\\\noalign{\medskip}0\end {array} \right] \qquad
|1\rangle =\left[ \begin {array}{c} 0\\\noalign{\medskip}1\end {array} \right]
$$
for the two spins and the operators
$$
  a=\left[ \begin {array}{cc} 0&1\\\noalign{\medskip}0&0\end {array} \right] \qquad
  a^\dagger = \left[ \begin {array}{cc} 0&0\\\noalign{\medskip}1&0\end {array} \right]
$$
which act on these states
$$
  a^\dagger |0\rangle = |1\rangle\qquad  a^\dagger |1\rangle = 0
$$
$$
  a |1\rangle = |0\rangle\qquad  a |0\rangle = 0\, .
$$
Here is an example of a state of the spin chain
$$
  |001011\rangle = |0\rangle\otimes|0\rangle\otimes|1\rangle\otimes|0\rangle\otimes|1\rangle\otimes|1\rangle
$$
for a system with 6 lattice sites. Label the sites starting from the left, as site 1, then site 2 and so on till we get
to the last site, which is site 6. The operator $a$ acting at the third site (for example) is
$$
  a_3 = 1\otimes 1\otimes a\otimes 1\otimes 1\otimes 1\, .
$$

We can then write $d^\dagger d$ as the following operator
\begin{eqnarray}
  d^\dagger d = \sum_p \sum_n a_n^\dagger a_p\, .
\end{eqnarray}
This is a long ranged spin chain. In terms of the Pauli matrices
$$
  \sigma^1=\left[ \begin {array}{cc} 0&1\\\noalign{\medskip}1&0\end {array} \right] \qquad
  \sigma^2=\left[ \begin {array}{cc} 0&-i\\\noalign{\medskip}i&0\end {array} \right]\qquad
  \sigma^3=\left[ \begin {array}{cc} 1&0\\\noalign{\medskip}0&-1\end {array} \right]
$$
we can write
$$
  a={1\over 2} (\sigma^1+i\sigma^2),\qquad a^\dagger ={1\over 2} (\sigma^1-i\sigma^2),\qquad a^\dagger a={1\over 2} (1-\sigma^3),
$$
\begin{equation}\label{}
 d^\dagger d = {1\over 4}\sum_p \sum_n (\sigma^1-i\sigma^2)_n (\sigma^1+i\sigma^2)_p
 =  -\sum_p {1\over 2} \sigma^3_p + {1\over 4}\sum_p \sum_n (\sigma^1_n\sigma^1_p + \sigma^2_n\sigma^2_p)\, .
\end{equation}

The total spins of the system are
$$
J^{3} =\sum_{p}\frac{1}{2}\sigma_{p}^{3},\;
J^{1}=\sum_{p}\frac{1}{2}\sigma_{p}^{1},\;
J^{2}=\sum_{p}\frac{1}{2}\sigma_{p}^{2}\, ,
$$
$$
\textbf{J}^{2}=J^{3}J^{3}+J^{1}J^{1}+J^{2}J^{2}\, .
$$
We use capital letters for operators and little letters for
eigenvalues. In terms of these total spins we have
$$
  d^\dagger d  =-J^{3}+(\textbf{J}^{2}-(J^{3})^{2})=
  \textbf{J}^{2}-J^{3}(J^{3}+1).
$$
Thus, eigenspaces of $d^\dagger d$ can be labeled by the
eigenvalues of $\textbf{J}^2$ and eigenvalues of $J^3$, and hence
the labels $R,(r,s)$ of the restricted Schur polynomial can be
traded for these eigenvalues. We will illustrate the connection
using a specific example and then state the general rule. Consider
the case of 8 spins. The impurities can be organized into any
irreducible representation corresponding to a Young diagram with 8
blocks and two rows. The possible irreducible representations are
{\small
$$
\yng(4,4)\quad \yng(5,3)\quad \yng(6,2)\quad \yng(7,1)\quad
\yng(8)
$$
}
and they have dimensions 14, 28, 20, 7 and 1 respectively. Coupling 8 spin-${1\over 2}$ particles using the usual rule for the addition
of angular momentum, we have
$$
{\bf {1\over 2}}\otimes {\bf {1\over 2}}\otimes {\bf {1\over 2}}\otimes {\bf {1\over 2}}\otimes {\bf {1\over 2}}\otimes
{\bf {1\over 2}}\otimes {\bf {1\over 2}}\otimes {\bf {1\over 2}} =
14\,{\bf 0}\oplus 28\, {\bf 1}\oplus 20\, {\bf 2}\oplus 7\, {\bf 3}\oplus {\bf 4}\, .
$$
Notice that the number of angular momentum multiplets matches the number of possible Young diagrams for the impurities and the
degeneracy of each multiplet matches the dimension of the $S_8$ irreducible representation associated to the Young diagram.
These connections are a consequence of the Schur-Weyl duality between the symmetric groups and the unitary groups.

The general rule is now clear: consider the restricted Schur
polynomial $\chi_{R,(r,s)}$. The $\textbf{J}^2 = j(j+1)$ quantum
tells you the shape of the Young diagram $s$ that organizes the
impurities. If there are $N_1$ boxes in the first row of $s$ and
$N_2$ boxes in the second, then $2j=N_1-N_2$. The $J^3$ eigenvalue
of the state is always a good quantum number, both in the basis we
start in where each spin has a sharp angular momentum or in the
basis where the states have a sharp total angular momentum. The
$j^3$ quantum number tells you how many impurities sit in the
first and second rows of $R$, that is, it tells you how many boxes
must be removed from each row of $R$ to obtain $r$. Denote the
number of boxes removed from the first row by $n_1$ and the number
of impurities in the second row by $n_2$. We have $2j^3=n_1-n_2$.
Here are some examples of the projection operators appearing in
the restricted Schur polynomials, written in terms of the $j,j^3$
states:
{\small
$$
P_{\tiny \yng(16,8), (\yng(12,4),\yng(4,4))}=\sum_{i=1}^{d_{\tiny {}_{\yng(4,4)}}} |j=0,j^3=0,i\rangle\langle j=0,j^3=0,i|
$$
$$
P_{\tiny \yng(16,8), (\yng(12,4),\yng(7,1))}=\sum_{i=1}^{d_{\tiny \yng(7,1)}} |j=3,j^3=0,i\rangle\langle j=3,j^3=0,i|
$$
$$
P_{\tiny \yng(16,8), (\yng(13,3),\yng(7,1))}=\sum_{i=1}^{d_{\tiny \yng(7,1)}} |j=3,j^3=-1,i\rangle\langle j=3,j^3=-1,i|
$$
$$
P_{\tiny \yng(16,8), (\yng(11,5),\yng(7,1))}=\sum_{i=1}^{d_{\tiny \yng(7,1)}} |j=3,j^3= 1,i\rangle\langle j=3,j^3=1,i|\, .
$$
}
In the above, $i$ labels all the states with the displayed
$(j,j^3)$ quantum numbers; it runs from 1 to the dimension of the
irreducible representation organizing the impurities. These
results are a general construction of the projection operators
used to define the restricted Schur polynomials dual to the two
AdS giant system.

When we evaluate the traces $(\ref{trace})$ we need to consider
the action of $\Gamma ((n,n+1))$. Towards this end, we again need
to switch to our previous convention of labeling $m+1$ boxes in
the Young diagram. Equivalently, we need add another spin site to
our chain. $\Gamma ((n,n+1))$ can then be taken to act on the
first and ($m+1$)th site of the spin chain\footnote{We could allow
$\Gamma ((n,n+1))$ to act on any of the first $m$ sites and the
($m+1$)th site of the spin chain without changing the final
result.}. Including this extra site our projectors become
$$
P_{R,(r,s)}=\sum_i |j,j^3,i\rangle\langle j,j^3,i|\otimes 1
$$
$$
=\sum_i |j,j^3,i\rangle\langle j,j^3,i|\otimes (|0\rangle\langle 0|+ |1\rangle\langle 1|)
$$
where the identity above sits in the ($m+1$)th slot. We will also
use the notation
$$
p_{R,(r,s)}=\sum_i |j,j^3,i\rangle\langle j,j^3,i|
$$
when the ($m+1$)th slot is not included.

In terms of the spin chain language, the intertwiners can be
written as
$$
I_1 = ({1\over 2} + {1\over 2} \sigma^3_{1}),  \qquad
I_2 = ({1\over 2} - {1\over 2} \sigma^3_{1}),
$$
$$
I_3 = a_1 ,\qquad I_4 = a_1^\dagger\,.
$$
When $\Gamma\left((n,n+1)\right)$ acts it does so by swapping the
first and ($m+1$)th spins, so that
\begin{eqnarray}
\nonumber
\Gamma \left((n,n+1)\right)I_1 &&= |0\rangle\langle 0|\otimes 1 \otimes \cdots \otimes 1 \otimes |0\rangle\langle 0|+
      |1\rangle\langle 0|\otimes 1 \otimes \cdots \otimes 1 \otimes |0\rangle\langle 1|\\
\nonumber
 &&= ({1\over 2} +{1\over 2} \sigma^3_1)({1\over 2} +{1\over 2} \sigma^3_{m+1})+a^\dagger_1a_{m+1}\, .
\end{eqnarray}
In exactly the same way
$$
I_1\Gamma \left((n,n+1)\right) = ({1\over 2} +{1\over 2} \sigma^3_1)({1\over 2} +{1\over 2} \sigma^3_{m+1})+a_1a^\dagger_{m+1}
$$
$$
\Gamma ((n,n+1)) I_1\Gamma ((n,n+1)) = ({1\over 2} +{1\over 2} \sigma^3_{m+1})\, .
$$
One of the terms contributing to the dilatation operator is
$$
{\rm Tr}(I_1 \big[ P_{R,(r,s)},\Gamma ((n,n+1))\big]I_1^T \big[P_{R,(r,s)},\Gamma ((n,n+1))\big])\, .
$$
This is a term that does not change the shape of the Young diagram labels of the restricted Schur polynomial.
Using the facts that $I_1=I_1^T$, $\Gamma \left((n,n+1)\right)^T =\Gamma \left((n,n+1)\right)$
and $P_{R,(r,s)}^T=P_{R,(r,s)}$ we can rewrite this term as
{\small
$$
2\left(
{\rm Tr}( I_1 P_{R,(r,s)}\Gamma ((n+1,n))I_1 P_{R,(r,s)}\Gamma ((n+1,n))) -
{\rm Tr}( I_1 P_{R,(r,s)} \Gamma ((n+1,n))I_1 \Gamma ((n+1,n))P_{R,(r,s)})\right)
$$
$$
= 2{\rm Tr}( P_{R,(r,s)}\big[({1\over 2} +{1\over 2} \sigma^3_1)({1\over 2} +{1\over 2} \sigma^3_{m+1})+a^\dagger_1a_{m+1}\big]
P_{R,(r,s)}\big[({1\over 2} +{1\over 2} \sigma^3_1)({1\over 2} +{1\over 2} \sigma^3_{m+1})+a^\dagger_1a_{m+1}\big])
$$
$$
-2{\rm Tr}( ({1\over 2} +{1\over 2} \sigma^3_{1}) P_{R,(r,s)} ({1\over 2} +{1\over 2} \sigma^3_{m+1}) P_{R,(r,s)})
$$
}
Tracing over the $m$th slot we easily find
$$
{\rm Tr}\left(P_{R,(r,s)}({1\over 2}+{1\over 2}\sigma^3_1)({1\over 2}+{1\over 2}\sigma^3_{m+1})P_{R,(r,s)}a_1^\dagger a_{m+1}\right)
={\rm Tr}\left(p_{R,(r,s)}({1\over 2}+{1\over 2}\sigma^3_1)p_{R,(r,s)}a_1^\dagger\right) \langle 0|1\rangle =0
$$
and
$$
{\rm Tr}\left(P_{R,(r,s)}a_1^\dagger a_{m+1}P_{R,(r,s)}a_1^\dagger a_{m+1}\right)
={\rm Tr}\left(p_{R,(r,s)}a_1^\dagger p_{R,(r,s)}a_1^\dagger\right)
{\rm Tr}\left(|0\rangle\langle 1|0\rangle\langle 1|\right)=0\, .
$$
The expression for the term in the dilatation operator that we are considering becomes
$$
{\rm Tr}(I_1 \big[ P_{R,(r,s)},\Gamma\left( (n,n+1\right) )\big]I_1^T \big[P_{R,(r,s)},\Gamma\left((n,n+1)\right)\big])
$$
$$
= 2{\rm Tr}\left( P_{R,(r,s)} ({1\over 2} +{1\over 2} \sigma^3_1)({1\over 2} +{1\over 2} \sigma^3_{m+1})
P_{R,(r,s)}({1\over 2} +{1\over 2} \sigma^3_1)({1\over 2} +{1\over 2} \sigma^3_{m+1})\right)
$$
$$
\left. -2{\rm Tr}( ({1\over 2} +{1\over 2} \sigma^3_{1}) P_{R,(r,s)} ({1\over 2} +{1\over 2} \sigma^3_{m+1}) P_{R,(r,s)}\right)\, .
$$
Expanding these expressions out and making use of the identities
$$
{\rm Tr}(P_{R,(r,s)}P_{R,(r,s)})= {\rm Tr}(P_{R,(r,s)}\sigma^3_{m+1}P_{R,(r,s)}\sigma^3_{m+1})
$$
$$
{\rm Tr}(P_{R,(r,s)}P_{R,(r,s)}\sigma^3_1)= {\rm Tr}(P_{R,(r,s)}\sigma^3_{m+1}P_{R,(r,s)}\sigma^3_1\sigma^3_{m+1})
$$
$$
{\rm Tr}(P_{R,(r,s)}\sigma^3_1 P_{R,(r,s)}\sigma^3_1)= {\rm Tr}(P_{R,(r,s)}\sigma^3_1\sigma^3_{m+1}P_{R,(r,s)}\sigma^3_1\sigma^3_{m+1})
$$
$$
{\rm Tr}(P_{R,(r,s)}P_{R,(r,s)}\sigma^3_{m+1})=0={\rm Tr}(P_{R,(r,s)}\sigma^3_1 P_{R,(r,s)}\sigma^3_{m+1})
$$
$$
{\rm Tr}(P_{R,(r,s)}P_{R,(r,s)}\sigma^3_1\sigma^3_{m+1})=0={\rm Tr}(P_{R,(r,s)}\sigma^3_1 P_{R,(r,s)}\sigma^3_1\sigma^3_{m+1})
$$
we obtain
$$
{\rm Tr}(I_1 \big[ P_{R,(r,s)},\Gamma ((n+1,n))\big]I_1^T \big[P_{R,(r,s)},\Gamma ((n+1,n))\big])
$$
$$
= {1\over 4}{\rm Tr}(P_{R,(r,s)}\sigma^3_1 P_{R,(r,s)}\sigma^3_1)-{1\over 4}{\rm Tr}(P_{R,(r,s)} )
$$
$$
= {\rm Tr}(P_{R,(r,s)}I_1 P_{R,(r,s)}I_1)-{\rm Tr}(P_{R,(r,s)}I_1 )\, .
$$
Using exactly the same types of arguments we find
{\small
$$
{\rm Tr}(I_1 \big[ P_{R,(r,s)},\Gamma ((n+1,n))\big]I_1^T \big[P_{S,(t,u)},\Gamma ((n+1,n))\big])
= {\rm Tr}(P_{R,(r,s)}I_1 P_{S,(t,u)}I_1)-{\rm Tr}(P_{R,(r,s)}I_1 )\delta_{R,(r,s)\,\, S,(t,u)}
$$

$$
{\rm Tr}(I_2 \big[ P_{R,(r,s)},\Gamma ((n+1,n))\big]I_2^T \big[P_{S,(t,u)},\Gamma ((n+1,n))\big])
= {\rm Tr}(P_{R,(r,s)}I_2 P_{S,(t,u)}I_2)-{\rm Tr}(P_{R,(r,s)}I_2 )\delta_{R,(r,s)\,\, S,(t,u)}
$$

$$
{\rm Tr}(I_3 \big[ P_{R,(r,s)},\Gamma ((n+1,n))\big]I_4^T \big[P_{S,(t,u)},\Gamma ((n+1,n))\big])
= {\rm Tr}(P_{R,(r,s)}I_2 P_{S,(t,u)}I_1)
$$

$$
{\rm Tr}(I_4 \big[ P_{R,(r,s)},\Gamma ((n+1,n))\big]I_3^T \big[P_{S,(t,u)},\Gamma ((n+1,n))\big])
= {\rm Tr}(P_{R,(r,s)}I_1 P_{S,(t,u)}I_2)\, .
$$
}

To evaluate these traces write each projector as ($i$ is a multiplicity label that runs from 1 to the
number of times the irreducible $SU(2)$ representation $j_1$ appears)
$$
  p_1=\sum_i |j_1,j^3_1 , i\rangle\langle j_1,j^3_1 , i |\, .
$$
Recall that we use a little letter $p$ for the projectors when the
($m+1$)th site is not included. In this last expression the
subscript 1 on the projector stands for $R,(r,s)$ and is a
notation which simplifies the equations dramatically. Introduce
the following states (the first state on the RHS is the ket
describing particle 1 in the lattice; the second ket is a good
total angular momentum state obtained by coupling the states of
the remaining $m-1$ particles):
$$
|\phi_1 ,i\rangle =|{1\over 2},{1\over 2}\, ;\, j_1-{1\over 2},j_1^3-{1\over 2}, i\rangle \langle{1\over 2},{1\over 2}\, ;\, j_1-{1\over 2},j_1^3-{1\over 2}, i|j_1,j_1^3\rangle
$$
$$
|\phi_2 ,i\rangle =|{1\over 2},{1\over 2}\, ;\, j_1+{1\over 2},j_1^3-{1\over 2}, i\rangle \langle{1\over 2},{1\over 2}\, ;\, j_1+{1\over 2},j_1^3-{1\over 2}, i|j_1,j_1^3\rangle
$$
$$
|\phi_3 ,i\rangle =|{1\over 2},-{1\over 2}\, ;\, j_1-{1\over 2},j_1^3+{1\over 2}, i\rangle \langle{1\over 2},-{1\over 2}\, ;\, j_1-{1\over 2},j_1^3+{1\over 2}, i|j_1,j_1^3\rangle
$$
$$
|\phi_4 ,i\rangle =
|{1\over 2},-{1\over 2}\, ;\, j_1+{1\over 2},j_1^3+{1\over 2}, i\rangle \langle{1\over 2},-{1\over 2}\, ;\, j_1+{1\over 2},j_1^3+{1\over 2}, i|j_1,j_1^3\rangle\, .
$$
Using these states the projector can be written as
$$
  p_1=\sum_i (|\phi_1 ,i\rangle +|\phi_3 ,i\rangle)(\langle \phi_1 ,i|+\langle \phi_3 ,i|)
     +\sum_i (|\phi_2 ,i\rangle +|\phi_4 ,i\rangle)(\langle \phi_2 ,i|+\langle \phi_4 ,i|)\, .
$$
These states are particularly convenient because the action of the intertwiners $I_1$ and $I_2$ on these states is very simple
$$ I_1 |\phi_1 ,i\rangle = |\phi_1 ,i\rangle,\qquad  I_1 |\phi_2 ,i\rangle = |\phi_2 ,i\rangle, $$
$$ I_1 |\phi_3 ,i\rangle = 0, \qquad  I_1 |\phi_4 ,i\rangle = 0, $$
$$ I_2 |\phi_1 ,i\rangle = 0,\qquad  I_2 |\phi_2 ,i\rangle = 0, $$
$$ I_2 |\phi_3 ,i\rangle = |\phi_3 ,i\rangle,\qquad  I_2 |\phi_4 ,i\rangle = |\phi_4 ,i\rangle\, . $$
We can now compute the traces we need
$$
  {\rm Tr}(P_1 I_1)=\sum_i\langle\phi_1,i|\phi_1,i\rangle + \sum_i\langle\phi_2,i|\phi_2,i\rangle
$$
$$
  = d_{R^{m+1}}\left[
    d_{s_1'}\left(\langle{1\over 2},{1\over 2}\, ;\, j_1-{1\over 2},j_1^3-{1\over 2} |j_1,j_1^3\rangle\right)^2 +
    d_{s_2'}\left(\langle{1\over 2},{1\over 2}\, ;\, j_1+{1\over 2},j_1^3-{1\over 2} |j_1,j_1^3\rangle\right)^2\right]\, .
$$
$s_1'$ is obtained by dropping a box from the first row of $s$ and $d_{s_1'}$ is the dimension of this irreducible representation.
$s_2'$ is obtained by dropping a box from the second row of $s$ and $d_{s_2'}$ is the dimension of this irreducible representation.
We have summed over $i$ to obtain the factors of $d_{s_1'}$ and $d_{s_2'}$. The overlaps $\langle\phi_1,i|\phi_1,i\rangle$
and $\langle\phi_2,i|\phi_2,i\rangle$ are independent of $i$ so that in the last line above we could sum over $i$ and
there is no further need for this index.
Recall the general expression for the Clebsch-Gordan coefficient
$$
\langle j_1,j_1^3\, ;\, j_2,j_2^3|j,j^3\rangle = \delta_{j^3,j^3_1+j^3_2}
\sqrt{(2j+1)(j+j_1-j_2)!(j-j_1+j_2)!(j_1+j_2-j)!\over (j_1+j_2+j+1)!}\times
$$
$$
\times
\sqrt{(j+j^3)!(j-j^3)!(j_1-j^3_1)!(j_1+j^3_1)!(j_2-j^3_2)!(j_2+j^3_2)!}
\times
$$
$$
\times\left(\sum_k
{(-1)^k\over k!(j_1+j_2-j-k)!(j_1-j^3_1-k)!(j_2+j^3_2-k)!(j-j_2+j_1^3+k)!(j-j_1-j^3_2+k)!}\right)^2
$$
where the sum runs over all values of $k$ for which the argument of each factorial is non-negative.
It is now straight forward to find
$$
  \langle{1\over 2},{1\over 2}\, ;\, j-{1\over 2},j^3-{1\over 2} |j,j^3\rangle =\sqrt{j+j^3+1\over
  2(j+1)},
\qquad
  \langle{1\over 2},{1\over 2}\, ;\, j+{1\over 2},j^3-{1\over 2} |j,j^3\rangle = \sqrt{j+j^3\over 2j},
$$
so that
$$
  {\rm Tr}(P_1 I_1)= 2d_{s_1'}d_{R^{m+1}} {j+j^3+1\over 2(j+1)} +  2d_{s_2'}d_{R^{m+1}} {j+j^3\over 2j}\, .
$$
where $R^{m+1}$ is a Young diagram of $S_{n-1}$, obtained from $R$
by removing $m+1$ boxes, and the factor of 2 comes from tracing
over the ($m+1$)th slot.
Now, since $I_1 + I_2$ is the identity we find
$$
 2 d_s d_{R^{m+1}} = {\rm Tr}(P_1) = {\rm Tr}(P_1 I_1) + {\rm Tr}(P_1 I_2) = d_s d_{R^{m+1}} + {\rm Tr}(P_1 I_2)
$$
which implies that
$$
  {\rm Tr}(P_1 I_2) = d_s d_{R^{m+1}} \, .
$$
Now consider the second type of term
$$
  {\rm Tr}(P_1 I_1 P_1 I_1)={\rm Tr}(I_1 P_1 I_1 P_1 I_1)
$$
where we used the fact that the trace is cyclic and $I_1$ is a projector so that $(I_1)^2=I_1$. This is a useful observation
because
$$
  I_1 P_1 I_1 P_1 I_1 = \sum_i \left(|\phi_1 ,i\rangle \langle \phi_1 ,i|+|\phi_2 ,i\rangle \langle \phi_2 ,i|\right)
                        \sum_j \left(|\phi_1 ,j\rangle \langle \phi_1 ,j|+|\phi_2 ,j\rangle \langle \phi_2 ,j|\right)
$$
It immediately follows that
$$
 {\rm Tr}(P_1 I_1 P_1 I_1)=2\left( d_{s_1'} \left({j+j^3+1\over 2(j+1)}\right)^2 +  d_{s_2'} \left({j+j^3\over 2j}\right)^2\right) d_{R^{m+1}} \, .
$$
Again using the fact that $I_1$ and $I_2$ sum to the identity, it is clear that
$$
 {\rm Tr}(P_1 I_2 P_1 I_1)= {\rm Tr}(P_1 I_1) - {\rm Tr}(P_1 I_1 P_1 I_1)
$$
$$
= 2d_{R^{m+1}}\left(d_{s_1'} \left({j+j^3+1\over 2(j+1)}\right) +  d_{s_2'} \left({j+j^3\over 2j}\right)
-d_{s_1'} \left({j+j^3+1\over 2(j+1)}\right)^2 -  d_{s_2'} \left({j+j^3\over 2j}\right)^2\right)\, .
$$
Finally,
$$
d_s d_{R^{m+1}} ={\rm Tr} (P_1)={\rm Tr} (P_1 P_1)
={\rm Tr} \left(P_1 (I_1+I_2)P_1 (I_1+I_2)\right)
$$
$$
={\rm Tr}(P_1 I_1 P_1 I_1)+2{\rm Tr}(P_1 I_2 P_1 I_1)+{\rm Tr}(P_1 I_2 P_1 I_2)\, .
$$
We can solve this last equation for ${\rm Tr}(P_1 I_2 P_1 I_2)$. This complete the evaluation
of the terms we were considering.

We now need to consider the case that the two projectors appearing in the trace have different labels.
There is only one term to compute because
$$
  {\rm Tr}(P_1 I_1 P_2 I_1) = {\rm Tr}(P_1 I_1 P_2 (I_1+I_2))-{\rm Tr}(P_1 I_1 P_2 I_2) = -{\rm Tr}(P_1 I_1 P_2 I_2)
$$
$$
 =-{\rm Tr}(P_1 (I_1 +I_2) P_2 I_2)+{\rm Tr}(P_1 I_2 P_2 I_2) = {\rm Tr}(P_1 I_2 P_2 I_2)
$$
The only time we get a non-zero result for this trace is when $j_2=j_1\pm 1$.
Without any loss of generality, consider the case that $j_2=j_1+1$.
The result for $j_2=j_1-1$ is obtained by swapping $1\leftrightarrow 2$.
We can write
$$
  p_2=\sum_i (|\psi_1 ,i\rangle +|\psi_3 ,i\rangle)(\langle \psi_1 ,i|+\langle \psi_3 ,i|)
     +\sum_i (|\psi_2 ,i\rangle +|\psi_4 ,i\rangle)(\langle \psi_2 ,i|+\langle \psi_4 ,i|)\, .
$$
where
$$
|\psi_1 ,i\rangle =|{1\over 2},{1\over 2}\, ;\, j_1+{1\over 2},j_1^3-{1\over 2}, i\rangle \langle{1\over 2},{1\over 2}\, ;\, j_1+{1\over 2},j_1^3-{1\over 2}, i|j_1+1,j_1^3\rangle
$$
$$
|\psi_2 ,i\rangle =|{1\over 2},{1\over 2}\, ;\, j_1+{3\over 2},j_1^3-{1\over 2}, i\rangle \langle{1\over 2},{1\over 2}\, ;\, j_1+{3\over 2},j_1^3-{1\over 2}, i|j_1+1,j_1^3\rangle
$$
$$
|\psi_3 ,i\rangle =|{1\over 2},-{1\over 2}\, ;\, j_1+{1\over 2},j_1^3+{1\over 2}, i\rangle \langle{1\over 2},-{1\over 2}\, ;\, j_1+{1\over 2},j_1^3+{1\over 2}, i|j_1+1,j_1^3\rangle
$$
$$
|\psi_4 ,i\rangle =|{1\over 2},-{1\over 2}\, ;\, j_1+{3\over 2},j_1^3+{1\over 2}, i\rangle \langle{1\over 2},-{1\over 2}\, ;\, j_1+{3\over 2},j_1^3+{1\over 2}, i|j_1+1,j_1^3\rangle
\, .
$$
We now find
$$
{\rm Tr}(P_1 I_1 P_2 I_1)= 2d_{R^{m+1}} {\rm Tr}\left[
\sum_i (|\psi_1 ,i\rangle \langle \psi_1 ,i| + |\psi_2 ,i\rangle \langle \psi_2 ,i|)
\sum_j (|\phi_1 ,j\rangle \langle \phi_1 ,j| + |\phi_2 ,j\rangle \langle \phi_2 ,j|)
\right]
$$
$$
= 2d_{R^{m+1}} \sum_{i,j}\langle\phi_2,j |\psi_1,i\rangle\langle\psi_1,i|\phi_2,j\rangle
$$
$$
= 2d_{R^{m+1}}d_{s_c'}(\langle{1\over 2},{1\over 2}\, ;\, j_1+{1\over 2},j_1^3-{1\over 2}, i|j_1,j_1^3\rangle)^2
          (\langle{1\over 2},{1\over 2}\, ;\, j_1+{1\over 2},j_1^3-{1\over 2}, i|j_1+1,j_1^3\rangle)^2
$$
$$
= 2d_{R^{m+1}}d_{s_c'}\left( {j_1+j^3_1 + 1\over 2(j_1 + 1)}\right)\left( {j_1-j^3_1 + 1\over 2(j_1 + 1)}\right) \, .
$$
If $s_1$ is the third label for $p_1$ and $s_2$ is the third label of $p_2$, then
by removing a box from $s_1$ we can get the same Young diagram as when we remove a box from $s_2$ - in the last line above
we have called this Young diagram which can be reached from either $s_1,s_2$ by removing a single box $s_c'$.

Putting things together we find (when $j=0$ the term ${(m+2)(j^3)^2\over j(j+1)}$ in round braces on the first line below
and the last term in the equation must be omitted)
\begin{eqnarray}\label{recursion_j3}
 && DO_{j,j^3}(b_{0},b_{1})=g_{YM}^2 \left[-{1\over 2}\left( m-{(m+2)(j^3)^2\over j(j+1)}\right)
\Delta O_{j,j^3}(b_{0},b_{1})\right.\nonumber \\
&& + \sqrt{(m + 2j + 4)(m - 2j)\over (2j + 1)(2j+3)} {(j+j^3 +
1)(j-j^3 + 1) \over 2(j + 1)}
 \Delta O_{j+1,j^3}(b_{0},b_{1})
\nonumber \\
&& \left. +\sqrt{(m + 2j + 2)(m - 2j +2)\over (2j + 1)(2j-1)}
{(j+j^3 )(j-j^3 ) \over 2 j} \Delta O_{j-1,j^3}(b_{0},b_{1})
\right]
\end{eqnarray}
where
\begin{eqnarray}
\Delta O(b_{0},b_{1}) &&=\sqrt{(N+b_{0})(N+b_{0}+b_{1})}(O(b_{0}+1,b_{1}-2)+O(b_{0}-1,b_{1}+2))\nonumber \\
 &&-(2N+2b_{0}+b_{1})O(b_{0},b_{1}).\label{}
\end{eqnarray}
This completes our evaluation of the dilatation operator for $m$
impurities.

\section{Construction of Sphere Giant Projectors and the Dilatation Operator}

The starting point of our AdS giant analysis was a polynomial
representation which was isomorphic to the Young diagrams. There
is a polynomial representation isomorphic to the sphere giants
too. For $m$ impurities, introduce the the $2m$ variables
$x_1,x_2,\cdots,x_m$ and $y_1,y_2,\cdots,y_m$. The $x$'s and $y$'s
are Grassman numbers and all these variables anticommute. The
$x$'s and $y$'s are associated to the two columns respectively.
Consider three impurities for illustration; in this case the
isomorphism is defined by
{\small
$$
\young(\,\,,\,\,,\,\,,\,\,,\,\,,\,,\,,3,2,1)\leftrightarrow y_1
y_2 y_3\quad
\young(\,\,,\,\,,\,\,,\,\,,\,1,\,,\,,\,,3,2)\leftrightarrow x_1
y_2 y_3\quad
\young(\,\,,\,\,,\,\,,\,\,,\,2,\,,\,,\,,3,1)\leftrightarrow x_2
y_1 y_3\quad
\young(\,\,,\,\,,\,\,,\,\,,\,3,\,,\,,\,,2,1)\leftrightarrow x_3
y_1 y_2
$$
$$
\young(\,\,,\,\,,\,\,,\,2,\,1,\,,\,,\,,\,,3)\leftrightarrow x_1
x_2 y_3\quad
\young(\,\,,\,\,,\,\,,\,3,\,1,\,,\,,\,,\,,2)\leftrightarrow x_1
x_3 y_2\quad
\young(\,\,,\,\,,\,\,,\,3,\,2,\,,\,,\,,\,,1)\leftrightarrow x_2
x_3 y_1\quad
\young(\,\,,\,\,,\,3,\,2,\,1,\,,\,,\,,\,,\,)\leftrightarrow x_1
x_2 x_3\, .
$$
}
The generalization to any $m$ is obvious. The polynomials are
ordered with (i) $x$'s to the left of the $y$'s and (ii) so that
their subscripts (within the $x$'s and $y$'s separately) increase.
The action of $S_n$ is to act on the subscripts without changing
the order of the variables.

The natural definition for $d$ is
\begin{equation}
 d=\sum_{i=1}^m y_i {\partial\over\partial x_i}\, .
\end{equation}
It is easy to check that $d$ does indeed commute with the
symmetric group.
Using the inner product: $\langle y_1y_2y_3,y_1y_2y_3\rangle = 1$
(so the inner product of same polynomials is 1 and of different
polynomials is 0) we find
\begin{equation}
 d^\dagger = \sum_{i=1}^m x_i{\partial\over\partial y_i}\, .
\end{equation}
The Casimir we want is
$$
  d^\dagger d =\sum_i x_i {\partial\over\partial x_i}-\sum_{i,j}x_i y_j {\partial\over\partial y_i}{\partial\over\partial x_j}\, .
$$

This has a nice expression in terms of the following spin model
\begin{eqnarray}
  d^\dagger d=\sum_{n,p}(-1)^{n-p}a_n^\dagger a_p = {1\over 4}\sum_{p,n} (-1)^{n+p}(\sigma^1_n\sigma^1_p +\sigma^2_n\sigma^2_p)
                                                 -{1\over 2}\sum_p\sigma_p^3\, .
\end{eqnarray}

If we further define
$$
  \tilde{\sigma}^i_n=(\sigma^3_n)^n \sigma^i_n(\sigma^3_n)^n
$$
we map this into the AdS giant problem
$$
  d^\dagger d=\sum_{n,p}(-1)^{n-p}a_n^\dagger a_p = {1\over 4}\sum_{p,n} (\tilde{\sigma}^1_n\tilde{\sigma}^1_p
                         +\tilde{\sigma}^2_n\tilde{\sigma}^2_p)-{1\over 2}\sum_p\tilde{\sigma}_p^3
={\tilde {\textbf J}}^2 -\tilde{J}^3(\tilde{J}^3+1)\, .
$$

To get the sphere giant dilatation operator we will simply need to
rewrite the formulas for the intertwiners and $\Gamma \left(
(n,n+1) \right)$ in terms of the $\tilde{\sigma}^i$ and then
trace. The computations in this case are parallel to those in
section \ref{construction_A}. $I_1$ and $I_2$ are unchanged; $I_3$
and $I_4$ each pick up a minus sign. Since they always appear
together, this change does not affect the answer for the
dilatation operator at all. As far as $\Gamma \left( (n,n+1)
\right)$, it continues to swap the first and last slots. If $m$ is
even, $\Gamma \left( (n,n+1) \right)$ just picks up a minus sign
and hence, since $\Gamma \left( (n,n+1) \right)$ always appears
twice, the dilatation operator is unaffected. When $m$ is odd only
the elements of $\Gamma \left( (n,n+1) \right)$ which are on the
diagonal change sign. The intertwiners $I_1,I_2$ only pick up on
the diagonal elements and the intertwiners $I_3,I_4$ only pick up
off the diagonal elements. Thus, we only ever get products of off
diagonal elements of $\Gamma \left( (n,n+1) \right)$ with off
diagonal elements of $\Gamma \left( (n,n+1) \right)$ or products
of on the diagonal elements of $\Gamma \left( (n,n+1) \right)$
with on the diagonal elements of $\Gamma \left( (n,n+1) \right)$.
Thus, even for $m$ odd the dilatation operator is unaffected. This
proves that
$$
\Tr\Big(\Big[ \Gamma_R((n,n+1)),P_{R\to (r,s)}\Big]I_{R'\,
T'}\Big[\Gamma_T((n,n+1)),P_{T\to (t,u)}\Big]I_{T'\, R'}\Big)
$$
is the same for the AdS and sphere giant cases and proves our
previous observation that to get sphere from AdS we just replace
factors like $N+b_0+b_1\to N-b_0-b_1$. We can thus use the exactly
the same method as for AdS case. Consider the restricted Schur
polynomial $\chi_{R,(r,s)}$ where all Young diagrams has at most 2
columns. The $\tilde{\textbf J}^2 = j(j+1)$ quantum number again
tells you the shape of the Young diagram $s$ that organizes the
impurities. if there are $N_1$ boxes in the first column of $s$
and $N_2$ boxes in the second, then $2j=N_1-N_2$. The $j^3$
quantum number again tells you how many boxes are removed from the
first and second columns of $R$ to produce $r$. Denote the number
of boxes removed from the first column by $n_1$ and the number of
boxes removed from the second column by $n_2$. We have
$2j^3=n_1-n_2$. The dilatation operator is thus
\begin{eqnarray}
  DO_{j,j^3}(b_0,b_1)=g_{YM}^2 \left[-{1\over 2}\left( m-{(m+2)(j^3)^2\over j(j+1)}\right)
\Delta O_{j,j^3}(b_0,b_1)\right.\nonumber \\
  + \sqrt{(m + 2j + 4)(m - 2j)\over (2j + 1)(2j+3)} {(j+j^3 + 1)(j-j^3 + 1) \over 2(j + 1)}\Delta O_{j+1,j^3}(b_0,b_1)
\nonumber \\
  \left. + \sqrt{(m + 2j + 2)(m - 2j +2)\over (2j + 1)(2j-1)} {(j+j^3 )(j-j^3 ) \over 2 j}\Delta O_{j-1,j^3}(b_0,b_1)\right]
\end{eqnarray}
where
\begin{eqnarray}
\Delta O(b_0,b_1)&&
=\sqrt{(N-b_0-b_1)(N-b_0)}(O(b_0+1,b_1-2)+O(b_0-1,b_1+2))
\nonumber \\
 &&-(N-b_0) O(b_0,b_1)-(N-b_0-b_1) O(b_0,b_1)\, .
\end{eqnarray}

\section{Diagonalization of the Dilatation Operator}

{ In this section we introduce a transformation that reduces the
dilatation operator to a set of decoupled oscillators.
     The transformation is constructed by solving a three term recursion relation. This three term recursion relation
     is nothing but the recursion relation of certain Clebsch-Gordan coefficients. This allows the construction of
     operators with a good scaling dimension in terms of Hahn polynomials. In the limit that the number of impurities is
     very large, these wave functions become the wave functions of the two dimensional radial oscillator. }

{\vskip 0.2cm}

\subsection{$j^3=0$ case and the corresponding discrete wave equation}
\label{diagonal_j3zero}

For this case the action of the dilatation operator is
\begin{eqnarray}\label{recursion_j3_zero}
  DO_{j,0}(b_{0},b_{1})&&=g_{YM}^2 \left[-{1\over 2} m \Delta
O_{j,0}(b_{0},b_{1})
  + \sqrt{(m + 2j + 4)(m - 2j)\over (2j + 1)(2j+3)} {(j+ 1) \over 2}
 \Delta O_{j+1,0}(b_{0},b_{1}) \right.
\nonumber \\
 &&\left. + \sqrt{(m + 2j + 2)(m - 2j +2)\over (2j + 1)(2j-1)} {j\over 2}
\Delta O_{j-1,0}(b_{0},b_{1}) \right]
\end{eqnarray}
where
\begin{eqnarray}
\Delta O(b_{0},b_{1}) &&=\sqrt{(N+b_{0})(N+b_{0}+b_{1})}(O(b_{0}+1,b_{1}-2)+O(b_{0}-1,b_{1}+2))\nonumber \\
 &&-(2N+2b_{0}+b_{1})O(b_{0},b_{1}).\label{}
\end{eqnarray}

Make the following ansatz for the operators of good scaling
dimension
\begin{equation}
  \sum_{b_1}\, f(b_0,b_1)\, O_p(b_0,b_1)\,=\sum_{j,b_1}\, C_p(j)\, f(b_0,b_1)\,
  O_{j,0}(b_0,b_1)
\label{goodscalingansatz}
\end{equation}
and require that
\begin{eqnarray}
 -\alpha_p C_p(j) &&=-{1\over 2} m C_p(j)
 + \sqrt{(m + 2j + 4)(m - 2j)\over (2j + 1)(2j+3)} {(j+ 1) \over 2}C_p(j+1)\nonumber\\
\label{recursion} &&+ \sqrt{(m + 2j + 2)(m - 2j +2)\over (2j +
1)(2j-1)} {j\over 2}C_p(j-1) \, .
\end{eqnarray}
The dilatation operator does not change the number of $Z$s ($n$) or the number of $Y$s ($m$).
Since $n =2b_0+b_1$ we do not sum over $b_0$ and $b_1$ independently in (\ref{goodscalingansatz}).
Further, $b_1$ only takes odd or even values. The action of the dilatation operator then reduces to
\begin{eqnarray}
DO_p(b_0,b_1) =&& -\alpha_p g_{YM}^2[\sqrt{(N+b_{0})(N+b_{0}+b_{1})}(O_p (b_{0}+1,b_{1}-2)+O_p (b_{0}-1,b_{1}+2))\nonumber \\
 &&-(2N+2b_{0}+b_{1})O_p (b_{0},b_{1})].\label{oscillatoreqn}
\end{eqnarray}
Reducing the dilatation operator to a set of decoupled oscillators
amounts to determining the coefficients $C_n(j)$ and the values of
$\alpha_n$, by solving the recursion relation (\ref{recursion}).
Now, introduce the Clebsch-Gordan coefficients
$$
  C^{j_1\, , j_2\, , j_T}_{m_1\, , m_2\, , m_T}=\langle j_T,m_T |j_1,m_1,j_2,m_2\rangle
$$
which couples the state with two angular momenta $j_1$ and $j_2$
to a state with good total angular momentum $j_T$. The eigenvalue
of the 3-component of angular momentum is denoted in the above using an $m$.
The recursion relation for the Clebsch-Gordan coefficients is
$$
(2n-{m\over 2})C^{{m\over 4},{m\over 4},j}_{n-{m\over 4},{m\over 4}-n,0}
=\sqrt{(m-2j)(m+2j+4)\over (2j+1)(2j+3)}{j+1\over 2}C^{{m\over 4},{m\over 4},j+1}_{n-{m\over 4},{m\over 4}-n,0}
$$
$$
+\sqrt{(m-2j+2)(m+2j+2)\over (2j+1)(2j-1)}{j\over 2}C^{{m\over
4},{m\over 4},j-1}_{n-{m\over 4},{m\over 4}-n,0}.
$$

It is now clear that the solution to our recursion relation is $C_p(j)=(-1)^{-j} C^{{m\over 4},{m\over
4},j}_{p-{m\over 4},{m\over 4}-p,0}$ with $m$ the number of
impurities and
$$
-\alpha_p = -2p = 0,-2,-4,...,-m\, .
$$
From the rules for coupling angular momentum, for the $j^3=0$ case,
the range of $p$ is $0\le p\le\frac{m}{2}$. The energy spacing of
the oscillators in the 1-2 directions are $4{\alpha_p} g_{YM}^{2}=8p
g_{YM}^2$. Thus, in this $j^3=0$ case we obtain ${m\over 2}+1$
oscillators with frequencies $0,8g_{YM}^2,16g_{YM}^2,$ $...,
4mg_{YM}^2$. 

Our ``eigenfunctions'', the $C_p(j)$, are given by Clebsch-Gordan
coefficients, or equivalently by (dual) Hahn polynomials,
reviewed for example in \cite{finiteoscillator}. They can be written in terms
of the ${}_3F_2$ hypergeometric function as
\begin{eqnarray}
C_p(j)=(-1)^{{m\over 2}-p}\left({m\over
2}\right)!\sqrt{(2j+1)\over \left({m\over
2}-j\right)!\left({m\over 2}+j+1\right)!}
{}_3F_2\left({}^{-j,j+1,-p}_{-{m\over 2},1};1\right)
\end{eqnarray}
where the range of $j$ and $p$ are $0\le j\le\frac{m}{2}$, $0\le p\le\frac{m}{2}$.
The lowest energy eigenfunction corresponds to the BPS states.
We have fixed the number of $Y$s to be $m=\gamma N$ with $\gamma\ll
1$ and $b_1=O(N)$ so that
$$
{m\over b_1}\sim \gamma \ll 1\, .
$$
This is the condition needed to ensure the dramatic
simplifications of Young's orthogonal representation. We can think
in terms of a double scaling limit $m\to\infty$ and $b_1\sim
N\to\infty$ keeping $\gamma$ fixed and very small. In this
limit ${}_3F_2\left({}^{-j,j+1,-p}_{-{m\over 2},1};1\right)\to
L_p\left( {2j^2\over m}\right)$ with $L_p(\cdot )$ the Laguerre
polynomial, so that
$$
C_p (j)\to (-1)^{{m\over 2}-p}\sqrt{2\over
m}\sqrt{2j+1}e^{-{j^2\over m}}L_p\left({2j^2\over m}\right)\qquad
0\le j\le\frac{m}{2}\, .
$$
These coefficients become the wave function of the 2d radial Harmonic oscillator for the $s$-wave, i.e. without a centripetal force
\begin{equation} \label{wave_34}
{1\over 2}\left[ -{1\over r}{\partial\over\partial r}\left(
r{\partial\over\partial r}\right) +r^2\right] C_p(r) = (2p+1)
C_p(r)
\end{equation}
where $\small{r={\sqrt{2\over m}}j}$ ranges over $0\le r \le
{\sqrt{m \over 2}}$ and $C_p(r)=C_p(j)/\sqrt{r}$ in the continuum
limit. This is a radial oscillator in 2d (the 3-4 directions) with
the energy levels labeled by integer $p$. What is the dual
interpretation of these wave functions? To answer this we need to
know how $r$ maps into the spacetime description of the dual
physics. Since $r$ is a scaled version of $j$, different values of
the $r$ coordinate correspond to different Young diagrams
organizing the impurities. These Young diagrams all have two rows.
Based on our experience with the half BPS case, it seems natural
to associate each one of the rows with each one of the giant
gravitons. We will simply assume that this is the correct
interpretation; this point certainly deserves further study.
Recalling that $Y=\phi_3+i\phi_4$ we know that the number of $Y$s
in each operator tells us the angular momentum of the operator in
the 3-4 plane. Denote the two angular momenta of the two giant
gravitons $J^Y_1$ and $J^Y_2$. The relations between $j,m,J^Y_1$
and $J^Y_2$ are
$$
  J_1^Y+J_2^Y=m\, ,\qquad J_1^Y-J_2^Y=2j\, ,
$$
$$
  J_1^Y={m\over 2}+j\, ,\qquad J_2^Y={m\over 2}-j\, .
$$
$j$ (and hence $r$) is directly proportional to the difference in
angular momenta of the two states. Giving an angular momentum to
the gravitons will cause them to expand as a consequence of \cite{Myers:1999ps}. The separation between the two
gravitons in the 3-4 plane will (for separations small compared to
the radii of the giants) be directly proportional to the
difference in angular momenta of the two giants. Consequently, it
is natural to interpret $r$ as a coordinate for the radial
separation between the two giants in the 3-4 plane. How large is
this separation? Recall that the radius of a giant with angular
momentum $J$, in units of the AdS scale $R$ is $\sqrt{J\over N}R$.
The maximum value of the difference between the angular of the
giants, $j$, is bounded by $m$ so that the length between them is
always very small $\sim \sqrt{\gamma}R$. The separation between the giants in 3-4 direction 
is thus seen to be small in AdS scale units for these states, due to the $\gamma$ factor, but it
can be large in string scale units. In units of the
string length the separation between the two giants is
$$
\sqrt{{2 \over m}}{j\over N^{1\over 4}}(g_{YM}^2)^{1\over 4} l_s
$$
where in the above we used the approximation 
${({m \over 2 }+j)}^{1 \over 2}-{({m \over 2 }-j)}^{1 \over 2} \to \sqrt{{2 \over m}}j $.
This separation can be an $O(1)$ or very large length. Thus, we propose that
$r$ is a coordinate for the radial separation between the two
giants in the 3-4 plane, and the separation naturally ranges from the string
length up to small (but non-zero) distances in units of the AdS
scale. One still needs to solve the eigenvalue problem of
(\ref{oscillatoreqn}) that will determine a ``wave function'' for
the 1-2 plane  and we see that $Z=\phi_1+i\phi_2$ plane. Thus, our
operators are described by a wave function in four dimensional
space. It is rather natural to interpret this space as the 4d
K\"ahler base appearing in the construction of the 1/4 BPS
geometries in \cite{Chen:2007du,hai2}. We see very concretely the
emergence of local physics on the 4d space from the system of
Young diagrams labeling the restricted Schur polynomial. This is
strongly reminiscent of the 1/2 BPS case where the Schur
polynomials provide wave functions for fermions in a harmonic
oscillator and further, these wave functions very naturally
reproduce features of the geometries and the phase space \cite{Lin:2004nb}
(for a review see for example \cite{Koch:2010zza}).

For  
the two matrix model we are studying here it is not true that the two  
matrices $Z$,$Y$ commute. For this reason, we
can't simultaneously diagonalize them and there is no analog of
the eigenvalue basis that is so useful for the large $N$ dynamics
of single matrix models. For the subsystem describing the BPS states of the $Z$,$Y$ system
however \cite{Berenstein:2005aa} has deduced that we can indeed assume that $Z$,$Y$
commute in the interacting theory and hence there should be a
description in terms of eigenvalues. Further, the eigenvalue
dynamics is again supposed to be dynamics in an oscillator
potential with repulsions preventing the collision of eigenvalues.
We have described a part of the BPS sector (as well as non-BPS
operators) among the operators we have studied. In the case of a
single matrix it is possible to associate the rows of the Young
diagram labeling a Schur polynomial with the eigenvalues of the
matrix. This provides a connection between the eigenvalue
description and the Schur polynomial description for single matrix
models. Are the oscillators we find here a signal of simple
underlying eigenvalue dynamics? Is there a connection between the
Young diagram labels and eigenvalues?

We will now study in detail the state with $p=0$ so that
$\alpha_{0}=0$, i.e. we consider a BPS state. This is the ground
state of the wave equation (\ref{wave_34}). In this case, since
${}_3F_2({}^{-j,j+1,0}_{-{m\over 2},1};1)=1$,
\begin{eqnarray}
C_{0}(j)=(-1)^{m \over 2}(\frac{m}{2})!\sqrt{\frac{(2j+
1)}{(\frac{m}{2}-j)!(\frac{m}{2}+j+1)!}} ,\quad\;0\le
j\le\frac{m}{2}\, . \nonumber
\end{eqnarray}
If we now consider the large $m$ regime,
\begin{equation}
C_{0}(j)\rightarrow(\frac{m}{2})^{-\frac{1}{2}}(-1)^{m \over 2}
\sqrt{2j+1} e^{-\frac{j^{2}}{m}}, \qquad{\normalcolor
{\normalcolor \mbox{for}}}\,\,\, m\gg1,\,\,0\le j\ll
m\label{c_m/2_j} \label{largem2}
\end{equation}
This formula agrees beautifully with our numerically obtained
eigenfunctions - see figure 1 where we make the comparison.
\begin{figure}[h]
          \centering
          {\epsfig{file=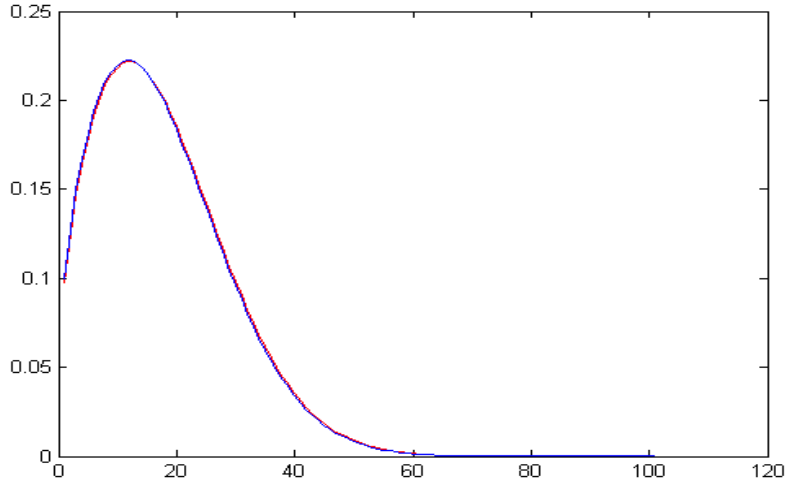,width=10.5cm,height=6.5cm}}
          \caption{A comparison between (\ref{largem2}) (plotted in red) and the numerically computed eigenfunction
                   for $m=600$. The agreement is clearly excellent.}
 \end{figure}

Another interesting eigenfunction to consider is
\begin{eqnarray}
C_{\frac{m}{2}}(j)=(\frac{m}{2})!\sqrt{\frac{(2j+1)}{(\frac{m}{2}-j)!(\frac{m}{2}+j+1)!}}
{} _{2}F_{1}(\begin{array}{c}
-j,j+1\\
1\end{array};1)\label{}  \nonumber
\end{eqnarray}
Using the identity
\begin{eqnarray}
_{2}F_{1}\Big(\begin{array}{c} -j,j+1\\
1\end{array};{1\over 2}{(1-z)}\Big)=P_{j}(z)\nonumber
\end{eqnarray}
where $P_{j}(z)$ is Legendre polynomial and $P_{j}(-1)=(-1)^{j}$,
we find
\begin{eqnarray}
C_{\frac{m}{2}}(j)=(\frac{m}{2})!(-1)^{j}\sqrt{\frac{(2j+
1)}{(\frac{m}{2}-j)!(\frac{m}{2}+j+1)!}} ,\quad\;0\le
j\le\frac{m}{2}\, .\nonumber
\end{eqnarray}
If we now consider the large $m$ regime,
\begin{equation}
C_{\frac{m}{2}}(j)\rightarrow(\frac{m}{2})^{-\frac{1}{2}} (-1)^{j}
\sqrt{2j+1} e^{-\frac{j^{2}}{m}}, \qquad{\normalcolor
{\normalcolor \mbox{for}}}\,\,\, m\gg1,\,\,0\le j\ll
m\label{c_m/2_j} \, .\label{largemlargeD2}
\end{equation}

This eigenfunction produces an oscillator with a level spacing of
$4mg_{YM}^2$. It is interesting to compare (\ref{largem2}) and
(\ref{largemlargeD2}): There are alternating minus signs, i.e.
$(-1)^j$ in (\ref{largemlargeD2}) for superposing
$j$-states, while there are same signs in (\ref{largem2}) for
superposing $j$-states; these different phases produce vastly different
eigenvalues for the two states. This again agrees very well with the
numerically computed eigenfunctions. We have compared the
numerically computed $C_{0}(j)$ and $C_{\frac{m}{2}}(j)$ in figure
2.
\begin{figure}[h]
          \centering
          {\epsfig{file=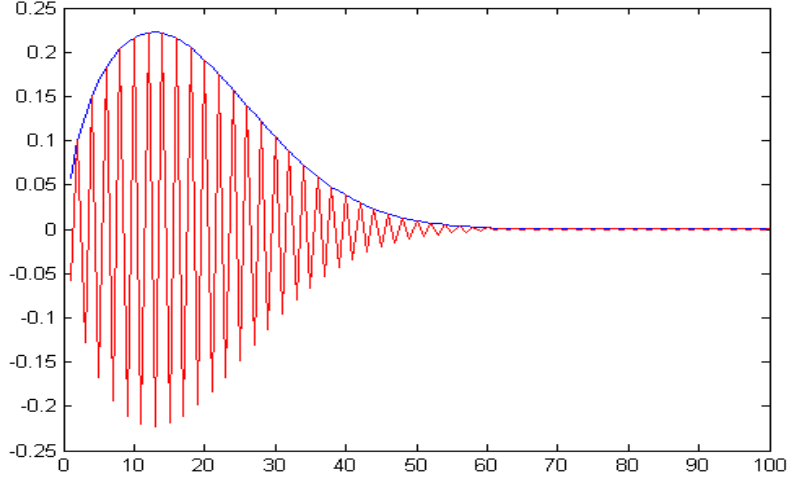,width=10.5cm,height=6.5cm}}
          \caption{A comparison between the numerically computed $C_{0}(j)$ (plotted in blue) and $C_{\frac{m}{2}}(j)$ (plotted in red) for $m=600$.
                   The two solutions are clearly identical up to the alternating signs in $C_{0}(j)$, exactly as predicted by the analytic expressions
                   for the eigenfunctions.}
 \end{figure}

\subsection{Arbitrary $j^3$ and the corresponding discrete wave equation}

For arbitrary $j^3$, we make the following ansatz for the
operators of good scaling dimension
$$
  \sum_{b_1}\, f(b_0,b_1)\, O_{p,j^3}(b_0,b_1)\,=\sum_{j,b_1}\, C_{p,j^3}(j)\, f(b_0,b_1)\,
  O_{j,j^3}(b_0,b_1)
$$

Repeating the arguments in the last subsection \ref{diagonal_j3zero},
we find that the $O_{p,j^3}(b_0,b_1)$'s satisfy the recursion equation
(\ref{oscillatoreqn}) with the prefactor given by
$-\alpha_{p,j^3}$. We also obtain the following recursion relation
\begin{eqnarray}
 && -\alpha_{p,j^3}C_{p,j^3}(j)=
  \sqrt{(m + 2j + 4)(m - 2j)\over (2j + 1)(2j+3)} {(j+j^3 + 1)(j-j^3 + 1) \over 2(j + 1)} C_{p,j^3}(j+1)
\nonumber \\
 && \sqrt{(m + 2j + 2)(m - 2j +2)\over (2j + 1)(2j-1)} {(j+j^3 )(j-j^3 ) \over 2 j}C_{p,j^3}(j-1)
  -{1\over 2}\left( m-{(m+2)(j^3)^2\over j(j+1)}\right)C_{p,j^3}(j)\, .\nonumber \\
\end{eqnarray}

This equation has a $j^3 \to -j^3$ symmetry, and we will solve for both the $j^3 \ge 0$ case
and the $j^3\le 0$ case. The range
for $j$ is
\begin{eqnarray}
0 \le |j^3|\le j\le\frac{m}{2}\, . \nonumber
\end{eqnarray}
The eigenfunctions $C_{p,j^3}(j)$ which solve the recursion relation are easily
expressed in terms of the hypergeometric functions ${}_3F_2$,
\begin{equation}
C_{p,j^3}(j)=(-1)^{{m\over 2}-p}\left({m\over
2}\right)!\sqrt{(2j+1)\over \left({m\over
2}-j\right)!\left({m\over 2}+j+1\right)!}
{}_3F_2\left({}^{|j^3|-j,j+|j^3|+1,-p}_{|j^3|-{m\over 2},1};1\right)
\label{universalefuncs}
\end{equation}
where the range of $j$ and $p$ are $|j^3| \le j\le\frac{m}{2}$, $0\le p\le\frac{m}{2}-|j^3|$, and
the associated eigenvalues are
$$ -\alpha_{p,j^3} = -2p = 0,-2,-4,...,-(m-2|j^3|)\, .
$$
This implies an energy spacing for the oscillators in 1-2
direction of $4{\alpha_p} g_{YM}^{2}=8p g_{YM}^2$. Thus, for a
given value of $j^3$ we obtain ${m\over 2}-|j^3|+1$ oscillators
with frequencies $0,8g_{YM}^2,16g_{YM}^2,$ $...,
(4m-8|j^3|)g_{YM}^2$. The lowest energy eigenfunction corresponds
to the BPS states. The eigenfunctions (\ref{universalefuncs}) are
related to the Hahn polynomials which are defined by
$$
Q_n(x;\alpha,\beta,N)={}_3F_2\left(^{-n,n+\alpha+\beta+1,-x}_{\alpha+1,-N};
1\right),\qquad n=0,1,2,\cdots, N.
$$
It is a well known fact that the Hahn polynomials are closely
related to the Clebsch-Gordan coefficients of $SU(2)$ \cite{HP}.

One can take a double scaling limit, so that
$m\to\infty$, $b_1 \to\infty$, ${m\over b_1}\sim\gamma \ll 1$, and
$\sqrt{2\over m}j$ becomes a continuous variable. In this limit,
${}_3F_2\left({}^{|j^3|-j,j+|j^3|+1,-p}_{|j^3|-{m\over
2},1};1\right)\to L_p( {2(j^2-|j^3|^2)\over m})$ with $L_p(\cdot
)$ the Laguerre polynomial. If we define the continuous variables
$\small{u_0=\sqrt{2\over m}j^3}$, $\small{r=\sqrt{2\over m}j}$,
$r^2=u^2+u_0^2$, and $j\ge |j^3|$, then we obtain the following
differential equation as the continuous
approximation of the difference equation
\begin{eqnarray}
{1\over 2}\left[ -{1\over u}{\partial\over\partial u}\left(
u{\partial\over\partial u}\right) +u^2\right] C_{p,u_0}(u) =
(2p+1) C_{p,u_0}(u)\, . \nonumber
\end{eqnarray}
In this equation $C_{p,u_0}=C_{p,j^3}(j)/\sqrt{r}$. Note that one can always
introduce an overall (normalization) factor in $C_{p,u_0}$. This
limit captures the continuous limit that we took when replacing the
hypergeometric function ${}_3F_2$ with the Laguerre polynomial $L_p(
{2(j^2-|j^3|^2)\over m})$,
$$
C_{p,j^3} (j)\to (-1)^{{m\over 2}-p}\sqrt{2\over
m}\sqrt{2j+1}e^{-{(j^2-|j^3|^2)\over m}}L_p \Big(
{2(j^2-|j^3|^2)\over m}\Big)
$$
where we have introduced an overall normalization factor.

For fixed $m$, $j^3$ ranges from $0\le |j^3|\le\frac{m}{2}$.
For the extreme case $j^3=\pm \frac{m}{2}$, there is only one
oscillator with zero frequency. All states of the corresponding oscillator are BPS states.
Another extreme case is $j^3= 0$. The eigenfunctions we have obtained in this section
nicely recover the results of subsection \ref{diagonal_j3zero} and we see that both the
$j^3=0$ and $j^3 \neq 0$ cases are captured by the eigenfunctions described in this section.

For a general $j^3$, the lowest energy state is given by the $p=0$
eigenfunction
\begin{equation}
C_{p=0,j^3}(j)=\left({m\over 2}\right)!(-1)^{{m\over 2}}
\sqrt{(2j+1)\over \left({m\over 2}-j\right)!\left({m\over
2}+j+1\right)!}. \label{C_j3_p0}
\end{equation}
The lowest energy eigenfunction corresponds to the BPS states.
The highest energy state is
\begin{equation}
C_{p=\frac{m}{2}-|j^3|,j^3}(j)=\left({m\over 2}\right)!
(-1)^{j}\left(^{j+|j^3|}_{j-|j^3|}\right) \sqrt{(2j+1)\over
\left({m\over 2}-j\right)!\left({m\over 2}+j+1\right)!},
\label{C_j3_pm2}
\end{equation}
where to obtain this simple result we have used the formulas
relating the hypergeometric function and the Jacobi polynomial
$P_{j-|j^3|}^{0,2|j^3|}(z)$:
$$
{}_2F_1({}^{|j^3|-j,j+|j^3|+1}_{1};{1\over
2}{(1-z)})=P_{j-|j^3|}^{0,2|j^3|}(z), \quad
P_{j-|j^3|}^{0,2|j^3|}(-1)=(-1)^{j-|j^3|}\left(^{j+|j^3|}_{j-|j^3|}\right).
$$
This eigenfunction (\ref{C_j3_pm2}) corresponds to an oscillator in the 1-2
directions with the largest energy spacing of $(4m-8|j^3|)g_{YM}^2$
in this specific $j^3$ sector.

It is interesting to compare (\ref{C_j3_p0}) and (\ref{C_j3_pm2}).
There are two main differences. One is the extra alternating signs
$(-1)^j$ in (\ref{C_j3_pm2}) for superposing $j$-states.
This is familiar from our $j^3=0$ results. Another difference is
the extra factor $\left(^{j+|j^3|}_{j-|j^3|}\right)$ which tends
to 1 when $j \to |j^3|$ and tends to $j^{(2|j^3|)} \over
(2|j^3|)!$ when $j \gg |j^3|$. Of course, in the special case
$j^3= 0$ we recover the eigenfunctions discussed in subsection
\ref{diagonal_j3zero}.

One can also look at the large $m$ regime of the eigenfunction
(\ref{C_j3_pm2}) which behaves as
\begin{eqnarray}
C_{p=\frac{m}{2}-|j^3|,j^3}(j)\rightarrow(\frac{m}{2})^{-\frac{1}{2}}
(-1)^{j} {\sqrt{2j+1}\over\Gamma(2|j^3|+1)}
{\Gamma(j+|j^3|+1)\over\Gamma(j-|j^3|+1)} e^{-\frac{j^{2}}{m}},
 \label{lstefunc}
\end{eqnarray}
for $m\gg 1,$ $|j^3|\le j\ll {m\over 2}$, and where
$\Gamma(\cdot)$ is the gamma function. Notice that in both the
limit $j \to |j^3|$ and the limit $j \gg |j^3|$ the eigenfunction
behaves nicely. The numerically generated eigenfunctions are
compared with the exact eigenfunctions in figure 3 below.
\begin{figure}[h]
          \centering
          {\epsfig{file=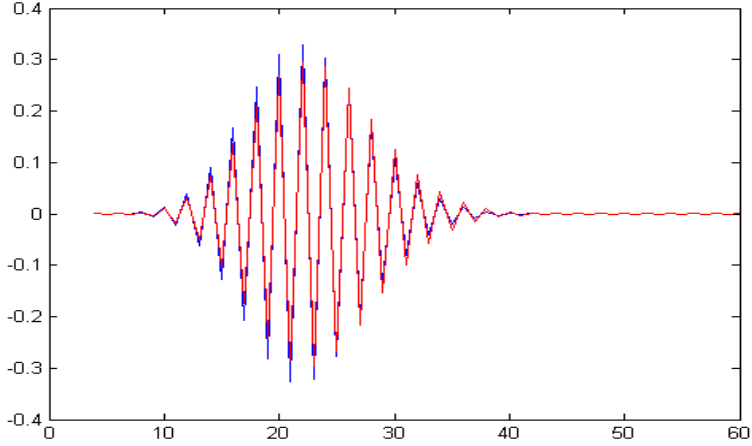,width=10.0cm,height=6cm}}
          \caption{A comparison between the numerically computed $C_{p=\frac{m}{2}-|j^3|,j^3}(j)$ (plotted in blue) and
                   the analytic formula (\ref{lstefunc}) (plotted in red) for $m=120$ and $j^3=4$. The agreement is
                   clearly excellent.}
 \end{figure}

\section{Discussion}
\label{discussion}

In summary we have found that: If the number of
impurities is even $=2n$ we obtain a set of oscillators with
frequency $\omega_i$ and degeneracy $d_i$ given by
$$\omega_i =8ig_{YM}^2,\qquad d_i=2(n-i)+1,\qquad i=0,1,...,n\, .$$
If the number of impurities is odd $=2n+1$
we obtain a set of oscillators with frequency $\omega_i$ and degeneracy $d_i$ given by
$$\omega_i =8ig_{YM}^2,\qquad d_i=2(n-i+1),\qquad i=0,1,...,n\, .$$
This is exactly the spectrum that was conjectured in
\cite{VinceKate}. In the present paper we have analytically
explained the degeneracies of the oscillators.

It is useful to review the salient features of our results. The
operators we consider, restricted Schur polynomials, are composed
of order $O(N)$ $Y$'s and $O(N)$ $Z$'s. Consequently, our large
$N$ spectra are obtained by summing both planar and nonplanar
diagrams. The nonplanar diagrams can not be neglected. Recall that
the restricted Schur polynomial has three labels $\chi_{R,(r,s)}$
with $R$ a Young diagram containing $n+m$ boxes, $r$ a Young
diagram containing $n$ boxes and $s$ a Young diagram containing
$m$ boxes. We have traded these labels for the integers $j$ (which
specifies $s$), $b_0,b_1$ (which specify $r$) and $j^3$ (which
specifies how $R$ and $r,s$ are related). By focusing on the
family of operators labeled using two column/row Young diagrams
and with fixed numbers of $Z$s and $Y$s we have simplified the
action of the dilation operator to two recursion relations, one in
the $b_0, b_1$ variables and one in the $j$ variables. A crucial
ingredient in our construction is the construction of the
projector $P$ used in defining the restricted Schur polynomial.
This construction is achieved by mapping each Young diagram $s$
onto a unique state of a spin chain with $m$ spin variables. The
computation of the projector is then reduced to the problem of
coupling the individual spins in the spin model to obtain a good
total spin.

The operators we have studied include both BPS and non-BPS states. The zero eigenvalue
states are the BPS states.
We find a discrete wave equation in the 3-4 directions (corresponding to $Y$) arising from
the recursion relation involving $j$. The associated ``energy levels'' are labeled by an
integer $p$ and set the parameter $\alpha_p$ which determines the frequencies of the
oscillators described by a second discrete wave equation in the 1-2 directions (corresponding
to $Z$). This second wave equation arises from the recursion relation involving $b_0,b_1$.
The associated energy levels are described in detail at the start of this section.
We have interpreted these states as the oscillations of the relative
positions of the two giant graviton branes. Our results
give a gauge theory description of the giant graviton brane
worldvolume physics.

Thanks to the cut off on the number of rows in the Young diagram, the sphere
giants form a cleanly decoupled sector. The same is not true of the AdS giants.
The fact that the AdS giants do not decouple at large $N$ as
cleanly as the sphere giants did, has not caused any problems.
Indeed, we have checked that all of the low lying energy levels
correspond to combinations of operators with all of their support
on operators labeled by Young diagrams with two long rows.

The spectra for the AdS giants and sphere giants are closely
related. This similarity between the spectra goes even further: we
have proved that the action of the dilatation operator on sphere
giants is related to its action on AdS giants upon making the
substitution $N+b\to N-b$. It would be nice to study this property
in more detail.

There are a number of new features of our results that deserve
comment. The spectrum we obtain is that of a set of oscillators
with frequency some multiple of $8g_{YM}^2$. It is somewhat
unusual to find a dependence on $g_{YM}^2$ itself at large $N$ -
we are used to the combination which defines the 't Hooft coupling
$\lambda= g_{YM}^2 N$. Our one loop results suggest that we can
hold $g_{YM}^2$ small but fixed. If we take the usual 't Hooft
limit, $g_{YM}^2\to 0$ and we obtain a continuous spectrum.
Our discrete spectra are obtained in a limit that is not the usual 't Hooft limit.
It would be very interesting to study the dilatation operator to
two loops and see if this dependence on $g_{YM}^2$, with no
dressing by factors of $N$ to some power, persists. Another
possibility is that we have an $SU(2)$ gauge symmetry for the
sector of the operators with two columns or two rows, and the
associated 't Hooft coupling is $2g_{YM}^2$.
The factor 2 may be understood as the number 
that multiplies the effective tension of the two coincident brane system.

The group $SU(2)$ has played a central role. It appears in at least three (apparently) unrelated ways
\begin{itemize}
\item The projectors $P_{R\to (r,s)}$ were written in terms of
$SU(2)$ Clebsch-Gordan coefficients. There is a natural action of
$SU(2)$ defined by our map to the spin chain (see sections 4, 5).
\item The eigenvalue problem for the $Z$ oscillator has been
solved in terms of symmetric Kravchuk polynomials. The symmetric
Kravchuk polynomials satisfy a difference equation that follows
from the raising and lowering relations between $SU(2)$ states
e.g. \cite{finiteoscillator}.
\item The eigenvalue problem for the
$Y$ oscillator has been solved in terms of Hahn polynomials.
Hahn polynomials are closely related to the Clebsch-Gordan
coefficients of $SU(2)$.
\end{itemize}

Since we deal with a two giant graviton system, the worldvolume
theory will have an $SU(2)$ gauge symmetry. Could this be related
to the $SU(2)$ groups we found above? It would be interesting to
study their relation.

The $SU(2)$s found above are not related to the global $SU(2)$ 
subgroup of the ${\cal R}$-symmetry, which rotates $Z$ and
$Y$ into each other. Indeed, the global $SU(2)$ rotates $Z$ and
$Y$ into each other so that their action mixes operators with
different values of $m$ and $n$. The $SU(2)$s found above all have
an action within a given $m,n$. It would be very interesting to
understand this further. This may well shed further light on the
integrability we have uncovered in this article.

{\vskip 1.0cm}

\noindent {\it Acknowledgements:} We would like to thank Tom Brown,
Vishnu Jejjala, Yusuke Kimura, Jurgis Pasukonis, Sanjaye Ramgoolam,
Peter Roenne and Dave Turton for correspondence and/or
discussions. This work is based upon research supported by the
South African Research Chairs Initiative of the Department of
Science and Technology and National Research Foundation. Any
opinion, findings and conclusions or recommendations expressed in
this material are those of the authors and therefore the NRF and
DST do not accept any liability with regard thereto. This work is
supported in part by Xunta de Galicia (Conselleria de Educacion
and grants PGIDIT10PXIB 206075PR and INCITE09 206121PR), by the
Spanish Consolider-Ingenio 2010 Program CPAN (CSD2007-00042), by
the Juan de la Cierva program of MCeI, and by the ME and Feder
(grant FPA2008-01838).

\appendix

\section{Dilatation Operator for AdS Giants with Two Impurities}
\label{A}

We are interested in the two AdS giant system for the case that
our operators have two impurities. In this case there are four
operators that can be produced once $r$ is given. These operators
are
{\small
$$
\chi_A(b_0,b_1)=\chi_{\tiny \yng(10,5)\, \yng(8,5)\, \yng(2)}(Z,Y)
\qquad
\chi_B(b_0,b_1)=\chi_{\tiny \yng(10,5)\, \yng(9,4)\, \yng(2)}(Z,Y)
$$
$$
\chi_C(b_0,b_1)=\chi_{\tiny \yng(10,5)\,\yng(9,4)\, \yng(1,1)}(Z,Y)
\qquad
\chi_D(b_0,b_1)=\chi_{\tiny \yng(10,5)\, \yng(10,3)\, \yng(2)}(Z,Y)
$$
}
The length of the rows in the first two labels are taken to infinity at large $N$. The
third label is exactly as shown even at large $N$. The number of boxes removed from each row of the first label
to obtain the second label are exactly as shown even at large $N$. The dilatation operator is

{\small
$$
D O_A (b_0,b_1)=
2g_{YM}^2(N+b_0+b_1+1)\left[
2{1\over (b_1+2)^2} O_A(b_0,b_1)\right.
$$
$$
+\left. {1\over (b_1+2)}\sqrt{b_1+1\over b_1+3} O_C(b_0-1,b_1+2)
-{b_1+4\over (b_1+2)^2}\sqrt{b_1+1\over b_1+3} O_B(b_0-1,b_1+2)\right]
$$
$$
+2g_{YM}^2\sqrt{(N+b_0+b_1+1)(N+b_0-1)}\left[ {b_1\over (b_1+2)^2}\sqrt{b_1+3\over b_1+1} O_B(b_0,b_1)\right.
$$
$$
\left.
-{1\over b_1+2}\sqrt{b_1+3\over b_1+1}O_C(b_0,b_1)
+2{1\over (b_1+2)^2} O_D(b_0-1,b_1+2)\right]
$$

$$ $$

$$
D O_D(b_0,b_1) =
2g_{YM}^2(N+b_0)\left[{b_1-2\over b_1^2}\sqrt{b_1+1\over b_1-1}O_B(b_0+1,b_1-2)\right.
$$
$$
-\left. {1\over b_1 }\sqrt{b_1+1\over b_1-1}O_C(b_0+1,b_1-2)
+2{1\over b_1^2} O_D(b_0,b_1)\right]
$$
$$
+2g_{YM}^2\sqrt{(N+b_0)(N+b_0+b_1)}\left[ {2\over b_1^2} O_A(b_0+1,b_1-2)\right.
$$
$$
\left. +{1\over b_1}\sqrt{b_1-1\over b_1+1}O_C(b_0,b_1)
-{b_1+2\over b_1^2}\sqrt{b_1-1\over b_1+1}O_B(b_0,b_1)\right]
$$

$$ $$

$$
D O_C(b_0,b_1)=
g_{YM}^2\sqrt{(N+b_0+1)(N+b_0+b_1+1)}\left[-{2\over b_1+2}\sqrt{b_1+3\over b_1+1} O_A(b_0,b_1)\right.
$$
$$
\left. - O_C(b_0-1,b_1+2)+{b_1+4 \over b_1+2} O_B(b_0-1,b_1+2)\right]
$$
$$
+g_{YM}^2 \sqrt{(N+b_0+b_1)(N+b_0)}\left[ {2\over b_1}\sqrt{b_1-1\over b_1+1} O_D(b_0,b_1)\right.
$$
$$
\left. -O_C(b_0+1,b_1-2)
+{b_1-2\over b_1}O_B(b_0+1,b_1-2)\right]
$$
$$
+g_{YM}^2\left( 2N+2b_0+b_1-3 \right) O_C(b_0,b_1)
$$
$$
+g_{YM}^2{(N+b_0)(4-4b_1-2b_1^2)-b_1^3-b_1^2+4b_1\over b_1(b_1+2)} O_B(b_0,b_1)
$$
$$
-2g_{YM}^2{(N+b_0-1)\over b_1+2}\sqrt{b_1+3\over b_1+1} O_D(b_0-1,b_1+2)
$$
$$
+2g_{YM}^2{(N+b_0+b_1)\over b_1}\sqrt{b_1-1\over b_1+1}O_A(b_0+1,b_1-2)
$$

$$ $$

$$
D O_B(b_0,b_1)=
g_{YM}^2\sqrt{(N+b_0-1)(N+b_0+b_1+1)}\left[
2{b_1\over (b_1+2)^2}\sqrt{b_1+3\over b_1+1}O_A(b_0,b_1)\right.
$$
$$
\left.
+{b_1\over b_1+2} O_C(b_0-1,b_1+2)
-{b_1(b_1+4)\over (b_1+2)^2} O_B(b_0-1,b_1+2)
\right]
$$
$$
+g_{YM}^2\sqrt{(N+b_0+b_1)(N+b_0)}\left[
-2{b_1+2 \over b_1^2}\sqrt{b_1-1\over b_1+1} O_D(b_0,b_1)\right.
$$
$$
\left.
+{b_1+2\over b_1} O_C(b_0+1,b_1-2)-{(b_1+2)(b_1-2)\over b_1^2} O_B(b_0+1,b_1-2)
\right]
$$
$$
+g_{YM}^2{(N+b_0)(4 -2b_1^2 -4b_1) - b_1^3 - b_1^2 + 4b_1 \over b_1(b_1+2)} O_C(b_0,b_1)
$$
$$
+g_{YM}^2{2(N+b_0)(b_1^4 + 4b_1^3 + 4b_1^2 - 8)+ b_1^5 + 5 b_1^4 + 8 b_1^3 - 16b_1\over b_1^2(b_1+2)^2} O_B(b_0,b_1)
$$
$$
+2g_{YM}^2{(N+b_0-1)b_1\over (b_1+2)^2}\sqrt{b_1+3\over b_1+1} O_D(b_0-1,b_1+2)
$$
$$
-2g_{YM}^2{(N+b_0+b_1)(b_1+2)\over b_1^2}\sqrt{b_1-1\over b_1+1} O_A(b_0+1,b_1-2)
$$
}

\section{Oscillators}
\label{Appendix_osc}

We look for operators with good scaling dimension that diagonalize the equation (\ref{oscillator01}). These take the form of
superpositions
\begin{eqnarray}
\sum_{b_{1}}f(b_{0},b_{1})O(b_{0},b_{1}) \nonumber
\end{eqnarray}
There are not independent sums over $b_0$ and $b_1$ because $n=2b_0+b_1$ is fixed.
Since
\begin{eqnarray}
  &&\sum_{b_{1}}f(b_{0},b_{1})DO(b_{0},b_{1})
= -\alpha
g_{YM}^{2}(\sum_{b_{1}}O(b_{0},b_{1})[\sqrt{(N+b_{0})(N+b_{0}+b_{1})}(f(b_{0}-1,b_{1}+2)\nonumber \\
&&+f(b_{0}+1,b_{1}-2))]-\sum_{b_{1}}O(b_{0},b_{1})(2N+2b_{0}+b_{1})f(b_{0},b_{1}))
= \sum_{b_{1}}\kappa f(b_{0},b_{1})O(b_{0},b_{1})\nonumber
\end{eqnarray}
where we have assumed $N+b_{0},b_{1} \gg 1$, the
$f(b_{0},b_{1})$ satisfy the recursion relation
\begin{eqnarray}
&&-\alpha g_{YM}^{2}[\sqrt{(N+b_{0})(N+b_{0}+b_{1})}(f(b_{0}-1,b_{1}+2)+f(b_{0}+1,b_{1}-2))\nonumber \\
&&-(2N+2b_{0}+b_{1})f(b_{0},b_{1})]=\kappa f(b_{0}, b_{1})
\label{Kp}
\end{eqnarray}
In the large $N+b_{0},b_{1}$ regime it is accurate to take a continuum limit of this recursion relation.
This gives a particularly simple description of the coefficients $f(b_{0},b_{1})$.
Towards this end, introduce the continuous variable $\rho=\frac{2b_{1}}{\sqrt{N+b_{0}}}$,
replace $f(b_0,b_1)$ with $f(\rho)$ and expand
$$
\sqrt{(N+b_0+b_1)(N+b_0)}=(N+b_0)\left( 1-{1\over 2}{b_1\over
N+b_0}-{1\over 8}{b_1^2\over (N+b_0)^2}+....\right)
$$
$$
f\left(\rho-{1\over\sqrt{N+b_0}}\right)=f(\rho)-{1\over\sqrt{N+b_0}}{\partial
f\over\partial \rho}+{1\over 2(N+b_0)}{\partial^2 f\over\partial
\rho^2}+...
$$
These expansions are only valid if $b_1\ll N+b_0$, which is
certainly not always the case. However, for eigenfunctions with
all of their support in the small $\rho$ region we do expect the
continuum limit of the recursion relation to give accurate answers. We find the recursion
relation becomes
\begin{equation}
(2\alpha g_{YM}^{2}){1 \over 2} \left[-{\partial^2 \over
\partial {\rho}^{2}}+\rho^{2}\right]f(\rho)=\kappa f(\rho)\label{reccc}
\end{equation}
which is a half of the harmonic oscillator with frequency $2\alpha
g_{YM}^{2}$. It is only half the oscillator because the lengths of the
rows (or columns) of the Young diagram are non-increasing. This implies that
$\rho\ge 0$, so that only half of the wavefunctions are selected (those
that vanish at $\rho=0$) and the energy spacing of the remaining oscillator states
is $4\alpha g_{YM}^{2}$. Clearly the description of the coefficients
$f(b_{0},b_{1})$ obtained by solving (\ref{reccc}) will be accurate for
the operators corresponding to the low lying oscillator eigenstates.

It is also possible to solve the recursion relation (\ref{Kp}) directly. Since we work in the large $N+b_{0},b_{1}$
regime, we can replace
(\ref{Kp}) by
\begin{eqnarray}
&&\kappa f(b_{0}, b_{1}) = -\alpha g_{YM}^{2}[\sqrt{(N+b_{0})(N+b_{0}+b_{1}+1)}f(b_{0}-1,b_{1}+2)\nonumber \\
&&+\sqrt{(N+b_{0}+1)(N+b_{0}+b_{1})}f(b_{0}+1,b_{1}-2)
-(2N+2b_{0}+b_{1})f(b_{1},b_{1})]
\nonumber
\end{eqnarray}
This recursion relation is solved by \cite{finiteoscillator}
\begin{equation}
f(b_0,b_1)={(-1)^n  ({1\over2})^{N+b_0+{b_1 \over
2}}}\sqrt{\left(^{2N+2b_0+b_1}_{N+b_0+b_1}\right)\left(^{2N+2b_0+b_1}_{\quad\quad
n}\right)}{}_2F_1({}^{-n,-(N+b_0+b_1)}_{-(2N+2b_0+b_1)};2)
\label{alllevels}
\end{equation}
where the hypergeometric function ${}_2F_1$ which appears defines the
symmetric Kravchuk polynomial $K_n(x,1/q,p)$
$$
 {}_2F_1\left({}^{-n,-x}_{-p};q\right)=K_n(x,1/q,p)\, .
$$
The corresponding eigenvalue is $\kappa=2n\alpha g_{YM}^2$. As for
the continuous solutions, because $b_1 \ge 0$ only half of the
wavefunctions are selected and the energy level spacing is again
$4\alpha g_{YM}^{2}$. The solutions (\ref{alllevels}) are accurate
for all levels of the oscillator. We have checked that the low
lying solutions of (\ref{alllevels}) are in excellent agreement
with the harmonic oscillator wave functions that vanish at the
origin. Note that for any finite value of $N$ the spectrum is
bounded. It is only in the $N\to\infty$ limit that the tower of
levels is infinite. In this limit, the harmonic oscillator wave
functions are an excellent description for any arbitrarily high
but finite energy level.

The analysis for the equation (\ref{oscillator02}) is similar once
we make the change $N+b_0\to N-b_0$. We have a similar recursion
relation for $f(b_{0}, b_{1})$ in this second case. By taking a
continuum limit, and using a continuous variable
$\tilde\rho=\frac{2b_{1}}{\sqrt{N-b_{0}}}$, we have
\begin{equation}
(2\alpha g_{YM}^{2}){1 \over 2} \left[-{\partial^2 \over
\partial {\tilde\rho}^{2}}+\tilde\rho^{2}\right]f(\tilde\rho)=\kappa f(\tilde\rho)\label{}
\end{equation}
which is again an oscillator with frequency $2\alpha g_{YM}^{2}$,
but since $\tilde\rho\ge0$, again, only half of the
wavefunctions are selected and the energy spacing is $4\alpha
g_{YM}^{2}$. Finally, the original recursion relation can again be
solved using symmetric Kravchuk polynomials.

\section{The Space $L(\Omega_p)$}
\label{rep}
In this Appendix we discuss the representation relevant for the problem of multiple AdS giants.
We highly recommend the article \cite{GelPair} for background material.
Consider $S_m$ the symmetric group on $m$-objects.
Then
$$
\Omega_p=S_m/S_{m-p}\times S_p
$$
is the space of all $p$ subsets of $\{1,2,...,m\}$. If $m=4$ then $\Omega_1=\{\{1\},\{2\},\{3\},\{4\}\}$ and
$\Omega_2=\{\{1,2\},\{1,3\},\{1,4\},\{2,3\},\{2,4\},\{3,4\}\}$ etc. You can identify a $p$ subset with a monomial.
For example, we'd identify $\{1,3\}$ with $x_1 x_3$ and $\{1,2,4\}$ with $x_1 x_2 x_4$. Thus, we can consider
$\Omega_p$ to be the space of distinct monomials with $p$ factors and no factor repeats. Ordering of the factors
is not important so that $x_1 x_2 x_4$ and $x_4 x_1 x_2$ are exactly the same element of $\Omega_3$. Our main interest
is in $L(\Omega_p)$ which is the space of complex valued functions on $\Omega_p$.
The symmetric group has a very natural action on $L(\Omega_p)$: we can define this action by defining it on each monomial.
The symmetric group acts by permuting the labels on the factors in the monomial. Thus, for example,
$$ (12)x_1 x_2 x_3 = x_1 x_2 x_3 \qquad
   (24)x_1 x_2 x_3 = x_1 x_4 x_3 \, .
$$
There is a natural inner product under which the monomials are orthonormal, so that, for example
$$
\langle x_1x_2x_3,x_1x_2x_3\rangle =1,\qquad \langle x_1x_2x_3,x_1x_2x_4\rangle =0= \langle x_1x_2x_3,x_1x_3x_4\rangle\, .
$$
$L(\Omega_p)$ furnishes a reducible representation of the symmetric group $S_m$. The relevance of $L(\Omega_p)$ for us here
is that the projectors acting in $L(\Omega_p)$ projecting onto an irreducible representation of $S_m$ are precisely the
projectors we need to define the restricted Schur polynomials. Consider the operator
\begin{equation}\label{}
d=\sum_{i=1}^m {\partial\over \partial x_i}\, .
\end{equation}
It maps from $L(\Omega_p)$ to $L(\Omega_{p-1})$. Further, it commutes with the action of $S_m$. Because
of this, elements of the kernel of $d$ form an invariant $S_m$ subspace. The intersection of the kernel
of $d$ and $L(\Omega_p)$ is called $S^{m-p,p}$ in \cite{GelPair} and it is proved that $S^{m-p,p}$ is an irreducible
representation of $S_m$.

An example will help to make this discussion concrete. Consider $S^{3,1}$ which is spanned by the polynomials
(this basis was found by writing the obvious polynomials linear in the $x_i$s that are annihilated by $d$ and
then using the Gram-Schmidt algorithm to get an orthonormal basis)
$$
\phi_1={x_1-x_2\over\sqrt{2}},\qquad
\phi_2={x_3-x_4\over\sqrt{2}},\qquad
\phi_3={x_1+x_2-x_3-x_4\over 2}\, .
$$
It is easy to check that
$$
(12)\phi_1=-\phi_1,\qquad
(12)\phi_2= \phi_2,\qquad
(12)\phi_3= \phi_3,
$$
$$
(23)\phi_1= {1\over 2}\phi_1-{1\over 2}\phi_2+{1\over\sqrt{2}}\phi_3,\qquad
(23)\phi_2=-{1\over 2}\phi_1+{1\over 2}\phi_2+{1\over\sqrt{2}}\phi_3,\qquad
(23)\phi_3= {1\over \sqrt{2}}\phi_1+{1\over \sqrt{2}}\phi_2,
$$
$$
(34)\phi_1= \phi_1,\qquad
(34)\phi_2=-\phi_2,\qquad
(34)\phi_3= \phi_3\, .
$$
Thus, we have the following group elements
$$
\Gamma\left( (12)\right)=
\left[\begin {array}{ccc} -1&0&0\\
\noalign{\medskip}0&1&0\\
\noalign{\medskip}0&0&1\end {array} \right]\qquad
\Gamma\left( (23)\right)=
\left[\begin {array}{ccc} {1\over 2}&-{1\over 2}&{1\over \sqrt{2}}\\
\noalign{\medskip}-{1\over 2}&{1\over 2}&{1\over \sqrt{2}}\\
\noalign{\medskip}{1\over \sqrt{2}}&{1\over \sqrt{2}}&0\end {array} \right]\qquad
\Gamma\left( (34)\right)=
\left[\begin {array}{ccc} 1&0&0\\
\noalign{\medskip}0&-1&0\\
\noalign{\medskip}0&0&1\end {array} \right]\, .
$$
Using these matrices it is possible to compute all elements of the group now, and then to compute characters. In this way,
it is a simple matter to identify this as the $\yng(3,1)$ irreducible representation.


\begin{thebibliography}{30}
\parskip-2pt


\bibitem{Maldacena:1997re}
  J.~M.~Maldacena,
  ``The large N limit of superconformal field theories and supergravity,''
  Adv.\ Theor.\ Math.\ Phys.\  {\bf 2}, 231 (1998)
  [Int.\ J.\ Theor.\ Phys.\  {\bf 38}, 1113 (1999)]
  [arXiv:hep-th/9711200];\\
  S.~S.~Gubser, I.~R.~Klebanov and A.~M.~Polyakov,
  ``Gauge theory correlators from non-critical string theory,''
  Phys.\ Lett.\ B {\bf 428}, 105 (1998)
  [arXiv:hep-th/9802109];\\
  E.~Witten,
  ``Anti-de Sitter space and holography,''
  Adv.\ Theor.\ Math.\ Phys.\  {\bf 2}, 253 (1998)
  [arXiv:hep-th/9802150].

\bibitem{Minahan:2002ve}
  J.~A.~Minahan and K.~Zarembo,
  ``The Bethe-ansatz for N = 4 super Yang-Mills,''
  JHEP {\bf 0303}, 013 (2003)
  [arXiv:hep-th/0212208].

\bibitem{Corley:2001zk}
  S.~Corley, A.~Jevicki and S.~Ramgoolam,
  ``Exact correlators of giant gravitons from dual N = 4 SYM theory,''
  Adv.\ Theor.\ Math.\ Phys.\  {\bf 5}, 809 (2002)
  [arXiv:hep-th/0111222],\\

\bibitem{McGreevy:2000cw}
  J.~McGreevy, L.~Susskind and N.~Toumbas,
  ``Invasion of the giant gravitons from anti-de Sitter space,''
  JHEP {\bf 0006}, 008 (2000)
  [arXiv:hep-th/0003075];\\
  M.~T.~Grisaru, R.~C.~Myers and O.~Tafjord,
  ``SUSY and Goliath,''
  JHEP {\bf 0008}, 040 (2000)
  [arXiv:hep-th/0008015];\\
  A.~Hashimoto, S.~Hirano and N.~Itzhaki,
  ``Large branes in AdS and their field theory dual,''
  JHEP {\bf 0008}, 051 (2000)
  [arXiv:hep-th/0008016].

\bibitem{Balasubramanian:2001nh}
  V.~Balasubramanian, M.~Berkooz, A.~Naqvi and M.~J.~Strassler,
  ``Giant gravitons in conformal field theory,''
  JHEP {\bf 0204}, 034 (2002)
  [arXiv:hep-th/0107119].

\bibitem{Berenstein:2004kk}
  D.~Berenstein,
  ``A toy model for the AdS/CFT correspondence,''
  JHEP {\bf 0407}, 018 (2004)
  [arXiv:hep-th/0403110].

\bibitem{Lin:2004nb}
  H.~Lin, O.~Lunin and J.~M.~Maldacena,
  ``Bubbling AdS space and 1/2 BPS geometries,''
  JHEP {\bf 0410}, 025 (2004)
  [arXiv:hep-th/0409174].

\bibitem{Balasubramanian:2005mg}
  V.~Balasubramanian, J.~de Boer, V.~Jejjala and J.~Simon,
  ``The library of Babel: On the origin of gravitational thermodynamics,''
  JHEP {\bf 0512}, 006 (2005)
  [arXiv:hep-th/0508023],\\
  V.~Balasubramanian, V.~Jejjala and J.~Simon,
  ``The library of Babel,''
  Int.\ J.\ Mod.\ Phys.\  D {\bf 14}, 2181 (2005)
  [arXiv:hep-th/0505123].

\bibitem{Grant:2005qc}
   L.~Grant, L.~Maoz, J.~Marsano, K.~Papadodimas and V.~S.~Rychkov,
  ``Minisuperspace quantization of 'bubbling AdS' and free fermion  droplets,''
  JHEP {\bf 0508}, 025 (2005)
  [arXiv:hep-th/0505079],\\
 G.~Mandal,
  ``Fermions from half-BPS supergravity,''
  JHEP {\bf 0508}, 052 (2005)
  [arXiv:hep-th/0502104].



\bibitem{Chen:2007gh}
  H.~Y.~Chen, D.~H.~Correa and G.~A.~Silva,
  ``Geometry and topology of bubble solutions from gauge theory,''
  Phys.\ Rev.\  D {\bf 76}, 026003 (2007)
  [arXiv:hep-th/0703068].

\bibitem{shahin}
  V.~Balasubramanian, J.~de Boer, V.~Jejjala and J.~Simon,
  ``Entropy of near-extremal black holes in AdS$_5$,''
  JHEP {\bf 0805}, 067 (2008)
  [arXiv:0707.3601 [hep-th]],\\
  R.~Fareghbal, C.~N.~Gowdigere, A.~E.~Mosaffa and M.~M.~Sheikh-Jabbari,
  ``Nearing Extremal Intersecting Giants and New Decoupled Sectors in N = 4
  SYM,''
  JHEP {\bf 0808}, 070 (2008)
  [arXiv:0801.4457 [hep-th]].

\bibitem{Koch:2008ah}
  R.~de Mello Koch,
  ``Geometries from Young Diagrams,''
  JHEP {\bf 0811}, 061 (2008)
  [arXiv:0806.0685 [hep-th]].

\bibitem{Koch:2008cm}
  R.~de Mello Koch, N.~Ives and M.~Stephanou,
  ``Correlators in Nontrivial Backgrounds,''
  Phys.\ Rev.\  D {\bf 79}, 026004 (2009)
  [arXiv:0810.4041 [hep-th]].

\bibitem{Koch:2009jc}
  R.~de Mello Koch, T.~K.~Dey, N.~Ives and M.~Stephanou,
  ``Correlators Of Operators with a Large R-charge,''
  arXiv:0905.2273 [hep-th].

\bibitem{us}
  R.~de Mello Koch, T.~K.~Dey, N.~Ives and M.~Stephanou,
  ``Hints of Integrability Beyond the Planar Limit,''
  JHEP {\bf 1001}, 014 (2010)
  [arXiv:0911.0967 [hep-th]].

\bibitem{hai1}
  H.~Lin, A.~Morisse and J.~P.~Shock,
  ``Strings on Bubbling Geometries,''
  JHEP {\bf 1006}, 055 (2010)
  [arXiv:1003.4190 [hep-th]].

\bibitem{hai2}
  H.~Lin,
  ``Studies on 1/4 BPS and 1/8 BPS geometries,''
  arXiv:1008.5307 [hep-th].

\bibitem{Balasubramanian:2004nb}
  V.~Balasubramanian, D.~Berenstein, B.~Feng and M.~x.~Huang,
  ``D-branes in Yang-Mills theory and emergent gauge symmetry,''
  JHEP {\bf 0503}, 006 (2005)
  [arXiv:hep-th/0411205].


\bibitem{de Mello Koch:2007uu}
  R.~de Mello Koch, J.~Smolic and M.~Smolic,
  ``Giant Gravitons - with Strings Attached (I),'' JHEP {\bf 0706}, 074 (2007),
  arXiv:hep-th/0701066.

\bibitem{de Mello Koch:2007uv}
  R.~de Mello Koch, J.~Smolic and M.~Smolic,
  ``Giant Gravitons - with Strings Attached (II),'' JHEP {\bf 0709} 049 (2007),
  arXiv:hep-th/0701067.

\bibitem{Kimura:2007wy}
  Y.~Kimura and S.~Ramgoolam,
  ``Branes, Anti-Branes and Brauer Algebras in Gauge-Gravity duality,''
  arXiv:0709.2158 [hep-th].

\bibitem{Bekker:2007ea}
  D.~Bekker, R.~de Mello Koch and M.~Stephanou,
  ``Giant Gravitons - with Strings Attached (III),''
  arXiv:0710.5372 [hep-th].

\bibitem{Brown:2007xh}
  T.~W.~Brown, P.~J.~Heslop and S.~Ramgoolam,
  ``Diagonal multi-matrix correlators and BPS operators in N=4 SYM,''
  arXiv:0711.0176 [hep-th].

\bibitem{Bhattacharyya:2008rb}
  R.~Bhattacharyya, S.~Collins and R.~d.~M.~Koch,
  ``Exact Multi-Matrix Correlators,''
  JHEP {\bf 0803}, 044 (2008)
  [arXiv:0801.2061 [hep-th]].

\bibitem{SS}
  S.~Corley, S.~Ramgoolam,
  ``Finite factorization equations and sum rules for BPS correlators in N=4 SYM theory,''
  Nucl.\ Phys.\ {\bf B641}, 131-187 (2002). [hep-th/0205221].

\bibitem{Brown:2008rr}
  T.~W.~Brown, P.~J.~Heslop and S.~Ramgoolam,
  ``Diagonal free field matrix correlators, global symmetries and giant
  gravitons,''
  arXiv:0806.1911 [hep-th].

\bibitem{Kimura:2008wy}
 Y.~Kimura and S.~Ramgoolam,
  ``Enhanced symmetries of gauge theory and resolving the spectrum of local
  operators,''
  Phys.\ Rev.\  D {\bf 78}, 126003 (2008)
  [arXiv:0807.3696 [hep-th]].

\bibitem{Kimura:2009wy}
 Y.~Kimura,
  ``Non-holomorphic multi-matrix gauge invariant operators based on Brauer
  algebra,''
  arXiv:0910.2170 [hep-th].

\bibitem{Ramgoolam:2008yr}
  S.~Ramgoolam,
  ``Schur-Weyl duality as an instrument of Gauge-String duality,''
  AIP Conf.\ Proc.\  {\bf 1031}, 255 (2008)
  [arXiv:0804.2764 [hep-th]].

\bibitem{Bhattacharyya:2008rc}
  R.~Bhattacharyya, R.~de Mello Koch and M.~Stephanou,
  ``Exact Multi-Restricted Schur Polynomial Correlators,''
  arXiv:0805.3025 [hep-th].

\bibitem{Brown:2008rs}
  T.~W.~Brown,
  ``Permutations and the Loop,''
  arXiv:0801.2094 [hep-th].

\bibitem{yusuke}
  Y.~Kimura,
  ``Quarter BPS classified by Brauer algebra,''
  JHEP {\bf 1005}, 103 (2010)
  [arXiv:1002.2424 [hep-th]].

\bibitem{Koch:2010gp}
  R.~d.~M.~Koch, G.~Mashile and N.~Park,
  ``Emergent Threebrane Lattices,''
  Phys.\ Rev.\  D {\bf 81}, 106009 (2010)
  [arXiv:1004.1108 [hep-th]].

\bibitem{Pasukonis:2010rv}
  J.~Pasukonis and S.~Ramgoolam,
  ``From counting to construction of BPS states in N=4 SYM,''
  arXiv:1010.1683 [hep-th].

\bibitem{VinceKate}
  V.~De~Comarmond, R.~de~Mello~Koch and K.~Jefferies,
  ``Surprisingly Simple Spectra,''
  [arXiv:1012.3884v1 [hep-th]].

\bibitem{Dunkl}
C. F. Dunkl, An addition theorem for Hahn polynomials: the spherical functions, SIAM J.
Math. Anal. 9 (1978), 627-637;\\
C. F. Dunkl, Spherical functions on compact groups and
applications to special functions, Symposia Mathematica 22 (1977),
145-161.

\bibitem{finiteoscillator}
N.M. Atakishiyev, G.S. Pogosyan and K.B. Wolf,
``Finite models of the Oscillator,''
Phys. Part. Nucl. 36, 521 (2005).

\bibitem{GelPair}
T. Ceccherini-Silberstein, F. Scarabotti and F. Tolli, Finite
Gelfand pairs and their applications to probability and
statistics, J. Math. Sci. N.Y., 141, no. 2 (2007), 1182-1229.

\bibitem{HP}
T.~H.~Koornwinder, ``Clebsch-Gordan coefficients for SU(2) and Hahn polynomials,'' Nieuw
Arch. Wisk. (3) 29, no. 2 (1981), 140-155,\\
A.~F.~Nikiforov, S.~K.~Suslov, ``Hahn polynomials and their
connection with Clebsch-Gordan coefficients of the group SU(2),''
Akad. Nauk SSSR Inst. Prikl. Mat. Preprint 1982, no. 83, 25p.

\bibitem{Chen:2007du}
A.~Donos,
  ``A description of 1/4 BPS configurations in minimal type IIB SUGRA,''
  Phys.\ Rev.\  D {\bf 75}, 025010 (2007)
  [arXiv:hep-th/0606199];\\
    B.~Chen {\it et al.},
  ``Bubbling AdS and droplet descriptions of BPS geometries in IIB
  supergravity,''
  JHEP {\bf 0710}, 003 (2007)
  [arXiv:0704.2233 [hep-th]];\\
    E.~Gava, G.~Milanesi, K.~S.~Narain and M.~O'Loughlin,
  ``1/8 BPS states in AdS/CFT,''
  JHEP {\bf 0705}, 030 (2007)
  [arXiv:hep-th/0611065];\\
  N.~Kim,
  ``AdS(3) solutions of IIB supergravity from D3-branes,''
  JHEP {\bf 0601}, 094 (2006)
  [arXiv:hep-th/0511029];\\
      O.~Lunin,
  ``Brane webs and 1/4-BPS geometries,''
  JHEP {\bf 0809}, 028 (2008)
  [arXiv:0802.0735 [hep-th]];\\
  J.~P.~Gauntlett, N.~Kim and D.~Waldram,
  ``Supersymmetric AdS(3), AdS(2) and bubble solutions,''
  JHEP {\bf 0704}, 005 (2007)
  [arXiv:hep-th/0612253];\\
  Z.~W.~Chong, H.~Lu and C.~N.~Pope,
  ``BPS geometries and AdS bubbles,''
  Phys.\ Lett.\  B {\bf 614}, 96 (2005)
  [arXiv:hep-th/0412221].


\bibitem{Biswas:2006tj}
  I.~Biswas, D.~Gaiotto, S.~Lahiri and S.~Minwalla,
  ``Supersymmetric states of N = 4 Yang-Mills from giant gravitons,''
  JHEP {\bf 0712}, 006 (2007)
  [arXiv:hep-th/0606087].

\bibitem{Mandal:2006tk}
  G.~Mandal and N.~V.~Suryanarayana,
  ``Counting 1/8-BPS dual-giants,''
  JHEP {\bf 0703}, 031 (2007)
  [arXiv:hep-th/0606088].

\bibitem{Beisert:2003tq}
  N.~Beisert, C.~Kristjansen and M.~Staudacher,
  ``The dilatation operator of N = 4 super Yang-Mills theory,''
  Nucl.\ Phys.\  B {\bf 664}, 131 (2003)
  [arXiv:hep-th/0303060].

\bibitem{Myers:1999ps}
  R.~C.~Myers,
  ``Dielectric-branes,''
  JHEP {\bf 9912}, 022 (1999)
  [arXiv:hep-th/9910053].

\bibitem{Koch:2010zza}
  R.~de~Mello~Koch and J.~Murugan,
  ``Emergent Spacetime,''
  arXiv:0911.4817 [hep-th].


\bibitem{Berenstein:2005aa}
  D.~Berenstein,
  ``Large N BPS states and emergent quantum gravity,''
  JHEP {\bf 0601}, 125 (2006)
  [arXiv:hep-th/0507203].


\end{thebibliography}
\end{document}